\documentclass[acmsmall,natbib=false]{acmart}

\usepackage{graphicx}
\usepackage{float}
\usepackage{tabularx}
\usepackage{booktabs}
\usepackage{algorithmic}
\usepackage{textcomp}
\usepackage{multirow}
\usepackage{threeparttable}
\usepackage{enumitem}
\usepackage[framemethod=tikz]{mdframed}
\usepackage{mfirstuc}
\usepackage{tikz}
\usepackage{tipa}
\usepackage{balance}
\usepackage{xtab}
\usepackage{mfirstuc}
\usepackage{todonotes}
\usepackage{soul}
\usepackage{colortbl}
\usepackage{xspace}
\usepackage{xcolor}
\usepackage{url}

\usepackage{caption}
\usepackage{subcaption}
\usepackage{wrapfig}
\usepackage[many]{tcolorbox}
\newtcolorbox{mybox}[1][]{
  breakable,
  fonttitle=\bfseries,
  bottomrule=0pt,
  toprule=0pt,
  leftrule=1pt,
  rightrule=1pt,
  titlerule=0pt,
  arc=0pt,
  outer arc=0pt,
  colframe=black,
}

\usepackage{listings}
\usepackage{color}

\definecolor{codebg}{rgb}{0.95,0.95,0.95}

\newboolean{showcomments}
\setboolean{showcomments}{true} %
\ifthenelse{\boolean{showcomments}}
{ \newcommand{\dnote}[2]{\textcolor{red}{
			\fbox{\bfseries\sffamily\scriptsize#1}
			{\small$\blacktriangleright$\textsf{\emph{#2}}$\blacktriangleleft$}}}}

\ifthenelse{\boolean{showcomments}}
    {
        \newcommand{\neil}[1]{\todo[inline,linecolor=yellow,backgroundcolor=yellow,bordercolor=black]{\textcolor{black}{\textbf{Neil says:}} #1}}
       \newcommand{\zane}[1]{\dnote{Zane}{#1}}
    }
    {
        \newcommand{\neil}[1]{}
        \newcommand{\zane}[1]{}
    }

\renewcommand{\theenumii}{\theenumi.\textbf{\arabic{enumii}}}

\RequirePackage[
  datamodel=acmdatamodel,
  style=acmauthoryear,
  ]{biblatex}

\setcopyright{acmlicensed}
\copyrightyear{2018}
\acmYear{2018}
\acmDOI{XXXXXXX.XXXXXXX}

\acmJournal{JACM}
\acmVolume{37}
\acmNumber{4}
\acmArticle{111}
\acmMonth{8}

\begin{document}

\title{The Nature of Technical Debt in Research Software}

\author{Neil A. Ernst}
\email{nernst@uvic.ca}
\orcid{0000-0001-5992-2366}
\affiliation{%
  \institution{University of Victoria}
  \city{Victoria}
  \state{BC}
  \country{Canada}
}
\author{Ahmed Musa Awon}
\email{its.ahmed.musa@gmail.com}
\affiliation{%
  \institution{University of Victoria}
  \city{Victoria}
  \state{BC}
  \country{Canada}
}
\author{Swapnil Hingmire}
\email{swapnilh@iitpkd.ac.in}
\affiliation{%
\institution{IIT Palakkad}
  \city{Palakkad}
  \country{India}
}
\author{Ze Shi Li}
\email{zeshili@ou.edu}
\affiliation{%
\institution{University of Oklahoma}
\city{Norman}
  \country{United States}
}

\begin{abstract}
Research software (also called scientific software) is essential for advancing scientific endeavours. Research software encapsulates complex algorithms and domain-specific knowledge and is a fundamental component of all science. A pervasive challenge in developing research software is technical debt, which can adversely affect reliability, maintainability, and scientific validity. Research software often relies on the initiative of the scientific community for maintenance, requiring diverse expertise in both scientific and software engineering domains. The extent and nature of technical debt in research software are little studied, in particular, what forms it takes, and what the science teams developing this software think about their technical debt. 
In this paper we describe our multi-method study examining technical debt in research software. We begin by examining instances of self-reported technical debt in research code, examining 28k code comments across nine research software projects. Then, building on our findings, we interview research software engineers and scientists about how this technical debt manifests itself in their experience, and what costs it has for research software and research outputs more generally.  
We identify nine types of self-admitted technical debt unique to research software, and four themes impacting this technical debt.
\end{abstract}


\received{20 February 2007}
\received[revised]{12 March 2009}
\received[accepted]{5 June 2009}

\maketitle

\section{Introduction}
 Research software, also known as scientific software, is essential for modern science, providing crucial tools for complex calculations, simulations, and data analysis across nearly all scientific disciplines. 
 It is integral to the formulation, testing, and validation of scientific hypotheses, with its accuracy and robustness directly impacting the validity of research findings. 
 This makes the development of research software a critical aspect of contemporary science, necessitating exceptional levels of precision and reliability \cite{Hook2009TestingFT,carver_software_2007}. 

 For example, in 2006, undetected software errors led to the retraction of five high-profile papers~\cite{Miller2006ASN}. These unintentional errors were not sophisticated: ``a homemade data-analysis program had flipped two columns of data, inverting the electron-density map'' yet went undetected for years, and were partially due to time pressure. This type of problem suggests that accumulated shortcuts---insufficient testing, inadequate code review, and missing validation of scientific assumptions---allowed the defects to persist. The significant ``interest'' paid in compromised accuracy and credibility illustrates how \textbf{technical debt in research software can have consequences beyond the codebase itself}, threatening the integrity of published scientific findings. Much like conventional software systems, research software requires active monitoring and management of technical debt.

 One challenge in research software development is that building research tools requires command of a diverse set of knowledge domains, each contributing to the overall precision and reliability of the software. According to \citet{kelly_scientific_2015}, these domains include 1) Real-World Knowledge, which involves understanding the scientific problems and data; 2) Theory-Based Knowledge, which encompasses the scientific principles and models underlying the software; 3) Software Knowledge, which involves programming and software engineering skills; 4) Execution Knowledge, related to running and testing the software; 5) Operational Knowledge, which pertains to the deployment and practical use of the software in real-world scenarios. 

The intersection of these domains poses significant challenges. To cross domains requires collaboration between scientists and software engineers (as well as other experts, e.g., in operating compute clusters), each bringing their expertise to ensure the software's quality, such as scientific accuracy and robustness. 
Such cross-domain challenges exist in most, if not all forms of software, as explored in research into social-technical congruence~\cite{Damian2013}, so the implications of this research are broad. We focus on research software as one software discipline with readily apparent Real-World and Theory-Based knowledge domains. Our goal is to explore the nature of technical debt in research software, including its extent, its characteristics, and how it is perceived by research software developers.
Technical debt provides a useful lens for understanding these cross-domain challenges because it captures not just individual defects, but the systemic accumulation of shortcuts and deferred decisions that erode software quality over time.

Understanding research software and technical debt is a multi-faceted, and socio-technical, research problem~\cite{Lamprecht2022WhatDW}. Software technical debt combines aspects of business and management, human motivations, and technical questions around mitigation and removal~\cite{ernst_technical_2021}. The study of such problems lends itself to a mixed-methods research approach (MMR)~\cite{storeyGuidelinesUsingMixed2024}. We approach the question using a quantitative study of self-admitted technical debt, and a qualitative study interviewing research software developers. 

Self-Admitted Technical Debt (SATD) provides a unique quantitative signal we can use to examine the challenges arising from the intersection of diverse knowledge domains necessary for developing robust software. SATD occurs when developers explicitly acknowledge, through code comments, that certain parts of the code are incomplete, need rework, contain errors, or are temporary solutions \cite{potdar_exploratory_2014}. In particular, these comments often reveal areas where the interplay between domain-specific knowledge and software engineering practices may compromise the reliability and correctness of the software. 

In the context of research software, SATD is particularly valuable as it highlights \textbf{cross-domain challenges faced by developers in integrating complex scientific theories and models with software implementations of those models}. These challenges are signaled by explicit indicators within SATD comments, pointing to underlying issues, complexities, or deficiencies in the code related to scientific accuracy, assumptions, or computational methods. Such signals provide direct insights into where and why the software might fail to meet scientific and engineering standards. Addressing these debts would not only enhance the maintainability of the software, but also ensure the accuracy and robustness of scientific computations, thereby safeguarding the integrity of research findings. 

\subsection{A Mixed Methods Approach}
\begin{figure}[ht]
    \centering
    \includegraphics[width=0.8\textwidth]{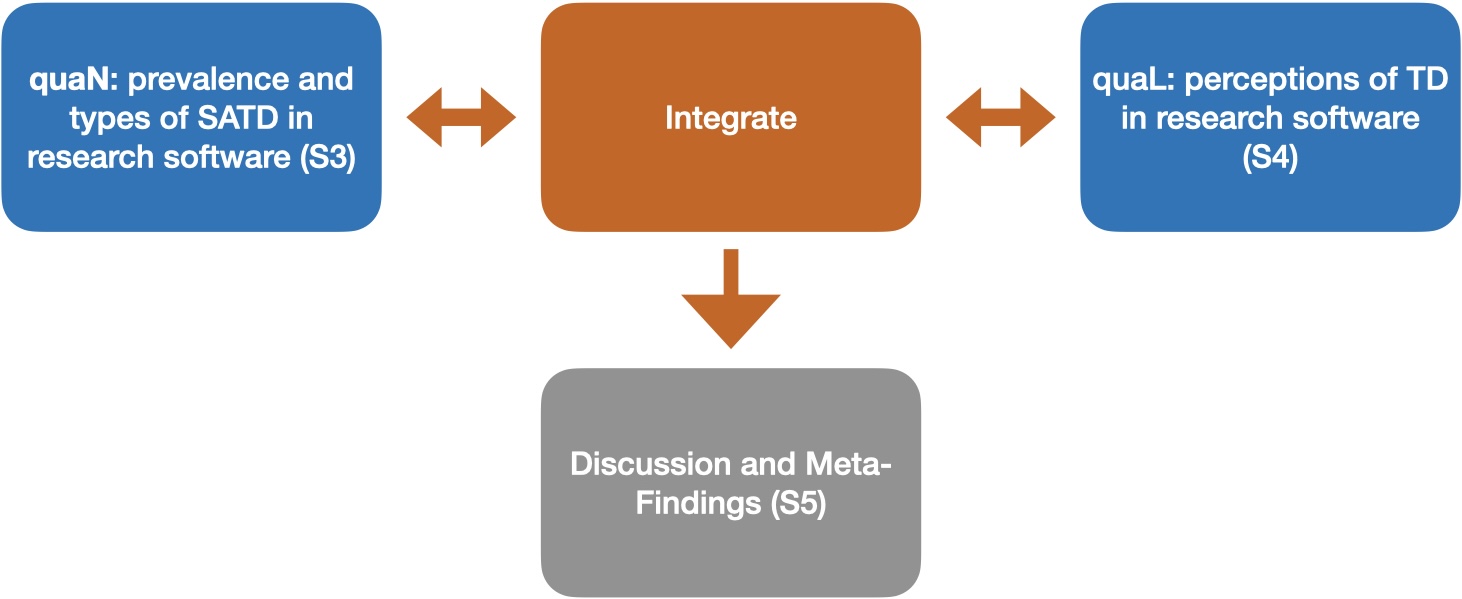}
    \caption{Overview of our Convergent Parallel Mixed Methods research approach. After ~\citet{storeyGuidelinesUsingMixed2024}}
    \Description{Flow diagram showing two parallel strands---quantitative mining and qualitative interviews---that are conducted concurrently, then merged and interpreted together.}
    \label{fig:mmr-flow}
\end{figure}

Our paper first introduces important background for our work, discussing technical debt and research software. 
To investigate the nature of technical debt in research software, we follow an \emph{Convergent Parallel mixed methods} approach~\cite{storeyGuidelinesUsingMixed2024} highlighted in Fig. \ref{fig:mmr-flow}. 

We gather quantitative insights (quaN) from large-scale analysis of research software code comments on self-admitted technical debt, reported in Section \ref{sec:quaN}. This study is motivated by our first research question:
\noindent\textbf{RQ1: How does SATD signal cross-domain challenges in Research Software?}
To answer this question, we collect a set of 28,680 self-admitted technical debt (SATD) comments from representative scientific projects, including high energy physics, astronomy, molecular dynamics, molecular biology, climate modeling, and applied mathematics. We then systematically label these comments using a combination of predefined and emergent labels. 

We then look to explain the results from that analysis using convergent analysis of our second study, an interview study of research software developers (Section \ref{sec:quaL}). This study builds on the insights we gain into research software and Scientific Debt to answer our second research question:
\noindent\textbf{RQ2: How do practitioners in research software projects perceive Technical Debt?}
To answer this question, we interviewed 11 contributors to long-lived research software projects about technical debt, conducted thematic analysis on their answers, and derived four themes to characterize the nature of technical debt.

We present each research question as two separate sections in this paper, each with self-contained details on method, analysis, and findings. A separate Discussion section (Section \ref{sec:discuss}) then integrates our findings from each method to synthesize insights. 

High-quality MMR studies follow four guiding principles: a) methodological rationale; b) novel integrated insights; c) procedural rigor, d) ethical research~\citet{storeyGuidelinesUsingMixed2024}. 

\textbf{Rationale}. We chose a mixed methods approach since the socio-technical nature of the problem, a complex interplay of human and organizational challenges occurring in the context of often highly complex and technical software projects, suggested that merely interviewing or mining would not be able to capture the complete picture. Interviewees may not be able to articulate the challenges with the code since it is so familiar to them; mining studies, especially for self-admitted technical debt, cannot capture the complete picture of technical debt in a project, just as it cannot for software bugs~\cite{arandaSecretLifeBugs2009}. In particular, our approach emphasizes Complementarity, Triangulation, and Explainability. 

\textbf{Integration} and \textbf{Rigor}. What did we gain from using a MMR approach? How well did we conduct the mixed studies? We comment on this in Section \ref{sec:discuss}.

\textbf{Ethical Research.} We sought and were granted review and approval for our study from our institution's IRB. In addition, our mining study followed ethical guidelines from \citet{goldEthicsMiningSoftware2022}.

\subsection{Contributions}
Emerging from our mixed methods study we contribute:
\begin{enumerate}
    \item Defining a new type of technical debt, termed \textbf{\textit{Scientific Debt}}, which highlights issues within the codebase acknowledged by contributors that could potentially compromise the validity, accuracy, and reliability of scientific results.
    \item A characterization of Scientific Debt across nine research software projects, including its distribution, sub-indicators (assumptions, translation challenges, missing edge cases, computational accuracy, and new scientific findings), and cross-project prevalence.
    \item A comprehensive dataset comprising 28,680 labeled SATD comments from various programming languages, marking the first instance of a SATD dataset that encompasses multiple programming languages.
    \item A repeatable coding guide to identify and categorize Scientific Debt.
    \item Empirically grounded themes and theory characterizing the nature, causes, and perception of technical debt in research software.
\end{enumerate}

\section{Background and Related Work}
What even \emph{is} research software? Taken broadly, the term suggests any use of software for research endeavours, and indeed software supports research in nearly every discipline, from humanities (text analysis) to engineering (building energy models) and social sciences (economic models). More recently, and more narrowly, research software has referred to any software written to support researchers, often by roles referred to as \emph{research software engineers}~\cite{pinto_how_2018}.

We focus on \emph{research} that pertains largely to natural science and engineering domains, e.g., math, physics, astronomy, climatology. Focusing on these domains emphasizes long-lived programs, which tended to originate in these fields before others, and tend to be large and complex. We comment on generality of our results later in the paper. We characterize \emph{software} more strictly. We deliberately exclude both software that plays an infrastructure role (operating systems, programming languages, message passing) and software that supports other software as helper libraries (such as matrix operations or supporting workflows). This definition mirrors the one chosen by \citet{kelly_scientific_2015} and used in the construction of a recent representative dataset, SciCat~\cite{MalviyaThakur2023SciCatAC}.

\subsection{Scientists and Research Software Development}
It is tempting to characterize research software developers as amateurs building tools as an ancillary task to their real interest, the domain science. To be sure, a large number of projects represent one-offs and paper/project-specific code that is never used by others (such as notebooks~\cite{pertsevaTheoryScientificProgramming2024}). However, many research software projects are long-lived and developed by large teams, such as the software used to process astronomical observations at the a radio telescope~\cite{ernstJSS}. In these projects, contributors are a mix of domain specialists (e.g., astronomy PhD students) and software experts (e.g., high performance computing programmers)~\cite{Glendenning2014TenTW}. It is these projects we focus on in this paper, and where technical debt is more likely to be an issue. 

Characterizing how research software is developed requires disentangling how the project operates, i.e., its software development process. Often scientists find that traditional software development methodologies, such as agile methodologies, do not fully address the unique needs of research software projects \cite{kelly_analysis_2011}. However, identifying a more suitable approach is challenging due to the lack of formal software development training among the scientists who typically lead these projects, and who often prioritize the correctness of scientific computations over software development processes. 

This prioritization results in an aversion to process-oriented approaches \cite{carver_software_2007}, leading them to adopt an "amethodical" approach \cite{Kelly2011ScientificST}. Many scientists learn software development through self-study and peer learning rather than formal education, which can result in poorly structured code and inadequate testing practices \cite{Wilson2006WheresTR,hannay_how_2009,pinto_how_2018}. Consequently, scientists tend to use programming languages, paradigms, and development environments they are familiar with, such as Fortran, which may not always align with best practices in software engineering \cite{kelly_assessing_2008, Segal2008ScientistsAS}.

Accuracy is a critical quality attribute for research software, as errors can severely compromise the validity and credibility of scientific findings \cite{meng_towards_2011}. However, achieving this accuracy is challenging for several reasons.
Research software needs to integrate knowledge across multiple domains, such as software knowledge, operations knowledge, and science/theory knowledge~\cite{kelly_scientific_2015}. 
These multiple domains are often the responsibility of different people, leading to fragmented team structures~\cite{arandaObservationsConwaysLaw2008}.
Gathering requirements for research software projects is particularly difficult due to the evolving and poorly defined nature of scientific inquiries \cite{carver_software_2007, kelly_software_2007}. 
This uncertainty complicates the process of ensuring that the software meets its intended requirements. 

Determining whether these requirements are met requires rigorous testing and validation, which are inherently complex in the context of research software. The intricate nature of scientific algorithms and the need to accurately model underlying scientific phenomena demand specialized testing methodologies tailored to the scientific domain \cite{carver_software_2007, kelly_assessing_2008}. 

The oracle problem, where it is difficult to determine the correct output due to complex computations, further complicates testing \cite{Hook2009TestingFT}. Moreover, these issues are exacerbated by the necessity of deep domain knowledge for both requirements gathering and validation \cite{Kelly2011SoftwareEF}. 
Even with this knowledge, scientists often build software to understand requirements better, and the lack of formal software development training can result in overlooked test cases. The high stakes involved are underscored by instances where software errors have led to the retraction of scientific papers, highlighting the critical importance of maintaining software accuracy \cite{Miller2006ASN}.

\subsubsection{Knowledge Domains}
Developing research software necessitates a profound integration of diverse knowledge domains to ensure accuracy and reliability. \citet{kelly_scientific_2015}'s \emph{knowledge acquisition model} highlights the importance of Real-World Knowledge, Theory-Based Knowledge, Software Knowledge, Execution Knowledge, and Operational Knowledge. For instance, constructing a climate model demands Real-World Knowledge about meteorological phenomena, such as understanding the dynamics of hurricanes. 

Theory-Based Knowledge is required to apply scientific principles and equations that govern atmospheric behavior. Software Knowledge is essential for programming these models efficiently. Execution Knowledge involves running and testing the software to ensure it produces accurate simulations, while Operational Knowledge pertains to deploying and using the software in real-world scenarios. This interdisciplinary approach is crucial for the successful development of research software, bridging theoretical concepts with practical implementations and ensuring the robustness of scientific findings.

Another primary challenge in the development of research software is ensuring long-term usability, as it must remain functional for many years and adapt to new scientific discoveries and technological advancements \cite{Kelly2009DeterminingFT,ernstJSS}. The "amethodical" development process, driven by the immediacy of publication and research needs, complicates long-term maintainability \cite{kelly_industrial_2013}. 
Research software tends to be highly customized for specific scientific questions, limiting its reuse and generalization \cite{Arnold2000DevelopingAA, Koteska2018}. This customization requires scientists to possess extensive software knowledge to maintain and update the software effectively, ensuring scalability and adaptability across different scientific domains.

\subsection{Technical Debt and Self-Admitted Technical Debt}
\textbf{Technical Debt (TD)} refers to the metaphorical ``borrowing" of shortcuts or suboptimal solutions in software development to achieve immediate gains, with the understanding that these decisions will incur ``interest" in the form of additional work required in the future if left unaddressed \cite{cunningham_wycash_1992}. 
The ``principal" in this context is the initial work deferred, which represents the tasks or improvements that were postponed. As time passes, interest accumulates, leading to increased maintenance costs and reduced system reliability. 

Since Ward Cunningham introduced the concept in 1992, it has evolved to encompass various software development trade-offs with long-term impacts \cite{kruchten_technical_2012}. \citet{lim_balancing_2012} found that while practitioners may not always recognize the term ``technical debt," they understand its implications, such as increased maintenance costs and reduced system reliability. TD includes different types such as code, design, architecture, and requirements debt, each posing unique challenges and requiring specific management strategies \cite{alves_towards_2014}. Design and architectural debts are particularly impactful, as they can lead to significant rework if not addressed early \cite{ernst_measure_2015}.

Vidoni and co-authors have studied technical debt in mathematical computing~\cite{Vidoni2022OnTD} and the R ecosystem~\cite{vidoni_self-admitted_2021}. They found that programmers in these domains were familiar with the ideas behind technical debt, if not the specific terminology, and that these programmers were usually quite deliberate and prudent in incurring technical debt. Vidoni coined the term \emph{Algorithm Debt}, which we use later in the paper, to capture the complex algorithms present in scientific code.

Graetsch et al.~\cite{Graetsch2025} conduct a study of data-intensive teams and determine new types of technical debt. The team they examine does not cross knowledge domains, focusing on data engineering, and does less software development. However, they note several data-focused types of technical debt, such as pipeline debt. We did not identify this as a form of SSATD, although most of our applications do feature data pipelines. Future exploration of the role of data-intensive technical debt in research software is important.

\textbf{Self-Admitted Technical Debt (SATD)} serves as a valuable signal to the various challenges and issues developers encounter in software development. 
Through intentional acknowledgment of technical debt in code comments, commit messages, or documentation, developers highlight areas of concern that need further attention. 
\citet{storey_todo_2008} found that annotations such as TODO and FIXME are widely used by developers to mark incomplete features, signal bugs, and indicate the need for refactoring. 
These annotations are prevalent, with 97\% of surveyed developers regularly employing them. 
Examples of SATD comments include statements like ``TODO - Move the next two subroutines to a new module called glad\_setup?".
or "FIXME(bja, $2016-10$) do these need to be strings, or can they be integer enumerations?"

\citet{potdar_exploratory_2014} formally defined SATD and identified that developers introduce it due to time pressure, the need for quick fixes, and the complexity of problems requiring temporary solutions. 
Their study revealed that SATD often remains unresolved for long periods, with only 26.3\% to 63.5\% of SATD being addressed across multiple releases. They identified 62 recurring patterns of SATD, which helps in recognizing the various forms of SATD across software projects.

Research software also contains SATD. \citet{sharma_self-admitted_2022} highlighted the gap in studying SATD within research software, particularly in dynamically-typed languages like R. They discovered that SATD in R packages often remains unaddressed, negatively impacting software quality and reliability. 
Similarly, \citet{liu_exploratory_2021} examined SATD in deep learning frameworks, finding that design debt is the most frequently introduced type, followed by requirement and algorithm debt. Their findings indicated that while requirement debt tends to be addressed promptly, documentation debt is frequently neglected. These studies underscore the potential of SATD to reveal underlying issues, such as bugs, incomplete features, and other challenges within a codebase, providing critical insights into the software development process. 

Research software is critical for reliable and accurate scientific results, yet the cross-domain nature of research software makes its construction difficult. While SATD exists in research software, it is not clear how its presence influences cross-domain challenges. We therefore investigate how technical debt influences research software. We begin with a quantitative exploration of SATD in research software, and then expand on those results with practitioner interviews.

\section{Study 1 - Quantitative Evaluation of SATD}
\label{sec:quaN}
Our first research question is \textbf{RQ1: How does SATD signal cross-domain challenges in Research Software?} In this section we describe the quantitative method we used to answer this question, and discuss findings. 

\subsection{Methodology}
\label{quaNmethod}
\subsubsection{Project Selection}

The selection of software projects for our analysis was guided by criteria aimed at capturing diverse practices and challenges across different scientific domains. We prioritized influential research software within their respective communities to ensure the relevance and impact of our findings.

Key selection criteria included:

\begin{itemize}
    \item \textbf{Renown}: Recognition and reputation within the scientific community.
    \item \textbf{Community Engagement}: Active, long-term discussions in project-specific forums, reflecting ongoing relevance and development.
    \item \textbf{Codebase Size}: Projects with at least 50,000 lines of code, as measured by the SLOCCount tool, ensuring medium to large scale.
    \item \textbf{Development Activity}: Regular updates and maintenance indicated by GitHub commit frequency.
    \item \textbf{Longevity}: Projects active for a minimum of 10 years, suggesting stability and enduring utility.
    \item \textbf{Popularity Metrics}: GitHub stars and forks, along with the number of dependent packages (DP) and dependent repositories (DR) for libraries.
\end{itemize}

Using these criteria, we purposively sampled nine exemplary projects that reflect the diversity and dynamism of advanced open-source software development in various scientific fields. Projects were selected to span diverse scientific domains (astronomy, high-energy physics, molecular biology, climate modeling, molecular dynamics, applied mathematics), programming languages (Python, C++, Fortran), and organizational structures. All nine projects met the minimum thresholds for codebase size, longevity, and development activity listed above. We did not systematically enumerate all candidate projects; rather, we drew on domain expertise and community recommendations to identify well-known, actively maintained projects in each field. Table \ref{tab:project_details} provides detailed descriptions of each selected project.

\begin{table*}[th]
    \centering
    \caption{Overview of case study projects. Note: Due to accessibility restrictions from GitLab, we could not extract the exact number of contributors and users for Athena.}
    \label{tab:project_details}
        \begin{tabular}{lp{1cm}ccccccc}
            \toprule
            \textbf{Name} & \textbf{Domain} & \textbf{Lang.} & \textbf{Contr.} & \textbf{SLOC} & \textbf{Stars} & \textbf{Forks} & \textbf{Age (Years)} \\
            \midrule
            Astropy & Astronomy & Python & 453 & 1,308,577 & 3.84K & 1.6K & 15 \\
            Athena & High Energy Physics & C++ & 100+ & 5,207,555 & -- & -- & 21 \\
            Biopython & Molecular Biology & Python & 331 & 620,437 & 3.61K & 1.63K & 27 \\
            CESM & Climate Model & Fortran & 134 & 2,799,805 & 265 & 154 & 43 \\
            GROMACS & Molecular Dynamics & C++ & 85 & 2,102,045 & 552 & 285 & 29 \\
            Moose & Physics & C++ & 221 & 847,602 & 1.5K & 979 & 18 \\
            Elmer & Applied Mathematics & Fortran/C++ & 45 & 954,420 & 1.1K & 292 & 12 \\
            Firedrake & Applied Mathematics & Python & 96 & 63,013 & 451 & 156 & 13 \\
            Root & High Energy Physics & C++ & 387 & 5,080,496 & 2.4K & 1.2K & 26 \\
            \bottomrule
        \end{tabular}
\end{table*}

\subsubsection{Extracting Comments} \label{secch3:extractComments}

Inspired by \citet{maldonado_using_2017} and \citet{liu_exploratory_2021}, we cloned the repositories of the selected projects and examined each file in the main or master branch. To handle the unique challenges of languages like C++ and Fortran, we created custom Python scripts using the GitPython library to explore the version control history.

For each file, we retrieved the initial commit using GitPython and extracted all comments from that point onward, recording metadata such as commit date, file name, and line number. To track comment removal, we recorded the date of the first commit in which the comment was no longer present. In line with \citet{freitas_comment_2012}, we treated multi-line comments as single, continuous comments in our analysis.

Errors could arise if files were renamed, relocated, or if SATD comments were modified. To minimize inaccuracies, we thoroughly reviewed comments and their context, cross-referencing commit histories and file changes to account for renaming or relocation and documenting instances where comments were altered rather than removed.

\subsubsection{Identifying SATD Instances}\label{secch3:identifySATD}

Manually identifying SATD from a large volume of comments is challenging and time-consuming. \citet{maldonado_using_2017} spent 185 hours classifying 62,556 comments. Our detection strategy draws inspiration from recent research by \citet{guo_how_2021}, confirming the effectiveness of keyword searches for identifying SATD instances. We used 64 established keywords identified by \citet{potdar_exploratory_2014}, known for flagging ``easy-to-find" instances of SATD, and an additional 597 keywords from \citet{sridharan_pentacet_2023} for detecting more subtle forms of technical debt.

All comments were preprocessed to normalize text, including tokenizing, converting text to lowercase, and stripping special characters. This ensured uniformity and improved keyword detection, recognizing variations like ``TODO," ``todo," or ``ToDo," and compound terms like ``pleasefixme."

A comprehensive keyword search reduced the number of comments to 39,697. However, this process also flagged non-relevant comments. For example, an Astropy comment:
\begin{quote}
    \textit{``Description: Generates a vector of length n containing the integers 0, ... , n-1 in random order. We \textbf{do not use} a new seed."}
\end{quote}
This was flagged due to the phrase \textit{do not use}. The second author (a graduate student with 2 years of experience in SATD research and software engineering) manually reviewed each of the 39,697 flagged comments, reading the comment text and its surrounding code context to determine whether the comment genuinely reflected a developer's acknowledgment of technical debt. Comments that matched keywords incidentally (as in the example above) were marked as non-relevant. This process took approximately \textbf{150} hours over \textbf{5} months, eliminating 11,017 non-relevant comments and resulting in 28,680 comments for further analysis. Our keyword-based approach prioritizes precision over recall: while we removed false positives through manual review, we did not conduct a systematic false negative analysis (i.e., sampling comments \emph{not} flagged by keywords to estimate how many SATD instances were missed). Prior work suggests keyword-based detection captures the majority of explicit SATD~\cite{potdar_exploratory_2014}, but more implicit or unusually phrased instances may be absent from our dataset.

\subsubsection{Categorizing SATD Instances}\label{secch3:categorizeSATD}

Given the dominance of Java in previous SATD research, automatic SATD categorization methods have primarily been developed for Java code comments, resulting in a high rate of false positives and limited detection of SATD types in our multi-language corpus. For example, \cite{li_automatic_2022} only detects four types of SATD. To identify a broader range of SATD types, including newer categories like ``On Hold Debt" and ``Algorithm Debt," and to uncover any new types of debt, we opted for a \emph{manual categorization} approach.

We employed the open card sorting technique~\cite{spencer_card_2009}, focusing on identifying the types of technical debt prevalent in the research software projects under study. 
Our categorization included Code Debt, Design Debt, Architectural Debt, Build Debt, Documentation Debt, Requirements Debt, Test Debt, and Defect Debt, as originally outlined in ~\cite{alves_towards_2014}. 
Additionally, we added categories for On Hold Debt and Algorithm Debt as described by \citet{maipradit_wait_2019} and  \citet{vidoni_self-admitted_2021} respectively. We frequently encountered blurred distinctions between different types of debt, aligning with findings from other studies \cite{maldonado_using_2017,li_automatic_2022}. In response, we developed operational definitions to clarify these distinctions, guiding our labeling of SATD comments. 

After reflecting on the data, we added a new category, termed \textbf{\textit{Scientific Debt}}, reflecting debt specific to the scientific nature of the software.
To mitigate personal bias in the manual classification of code comments, we implemented a systematic verification process. The second author randomly sampled a statistically representative subset of 1,000 SATD instances from the 28,680 identified instances, using a 95\% confidence level with a 10\% confidence interval. This sample was independently classified by the third author, a post-doctoral researcher. Inter-rater reliability was +0.79 (Cohen's kappa coefficient), indicating substantial inter-rater reliability. The kappa was computed on single-label classifications, following the approach of \citet{maldonado_using_2017}.

Subsequently, during disagreement resolution, we observed that many comments could reasonably fall under multiple categories; we then added those additional labels but did not conduct a second round of inter-rater agreement on the multi-label assignments. We included all relevant categories for each comment. For example, the comment \textit{TODO - Change which\_call to an integer? Modify for Glissade? (dissip has smaller vertical dimension)} was labeled as both Code Debt and Design Debt, because it indicates a need for changes in the code implementation (Code Debt) and suggests a potential alteration in the software's design or architecture to accommodate new requirements or improvements (Design Debt). This approach provides a nuanced and accurate representation of the self-admitted technical debt present in the code. 
The definitions of different categories of debt with examples are defined in Table \ref{tbl:codes}.

\begin{table*}[ht]
    \caption{SATD types used in the coding.}
    \label{tbl:codes}
    \resizebox{\textwidth}{!}{
    \begin{tabular}{@{}cp{2.5in}p{2.5in}c@{}}
        \toprule
        \textbf{Debt Type} & \textbf{Definition} & \textbf{Example} & \textbf{Source}\\ \midrule
        Architectural & Issues in project architecture, such as the violation of modularity, which can impact architectural requirements like performance and robustness. & \emph{CESM}: ``TODO - Move the higher-order stuff to the HO driver, leaving only the old Glide code." & \cite{alves_towards_2014}\\
        Build & Problems in dependency management and build processes, such as disorganized compile flags or problematic build targets, which complicate the build environment. & \emph{Root}: ``FIXME! This function is a workaround on OSX because it is impossible to link against libzmq.so" & \cite{alves_towards_2014}\\
        Code & Complex, obsolete, or redundant code that compromises code quality or fails to adhere to best coding practices. & \emph{Moose}: ``TODO: Rename this method to getName; the normal name (ID) should be getPath." & \cite{alves_towards_2014} \\ 
        Defect & Known bugs or issues within the software that are acknowledged but not yet corrected. & \emph{Elmer}: ``For some reason this is not always active. If not set, parallel interpolation could fail."  & \cite{alves_towards_2014}\\
        Design & Suboptimal design decisions that lead to inconsistent practices or insufficient modularization, complicating future modifications. & \emph{Biopython}: ``TODO - How to handle the version field? At the moment, the consumer will try to use this for the ID which isn't ideal for EMBL files."  & \cite{alves_towards_2014}\\
        Test & Issues related to the testing process, including costly or complex tests, lack of coverage, or inconsistent test results. & \emph{CESM}: ``NOTE (bja, 2018-03) ignoring for now... Not clear under what conditions the test is needed." \\
        Requirements & Unmet or partially implemented requirements, including non-functional ones (e.g., security, performance), affecting system functionality. & \emph{Astropy}: ``Binary FITS tables support TNULL *only* for integer data columns. TODO: Determine a schema for handling non-integer masked columns in FITS."  & \cite{alves_towards_2014}\\
        Docs & Missing, inadequate, or inaccurate documentation that fails to properly guide the user or developer. & \emph{Biopython}: ``TODO: add information about what is in the aligned species DNA before and after the immediately preceding 's' line."  & \cite{alves_towards_2014}\\
        Algorithm & Use of algorithms that are suboptimal or inadequately address the intended problem, resulting in inefficient performance or scalability issues. & \emph{Root}: ``TODO: We could optimize based on the knowledge that when splitting a failed partition into two, if one side checks out okay then the other must be a failure." & \cite{maipradit_wait_2019,vidoni_self-admitted_2021}\\
        On Hold & Development delays caused by waiting for other functionalities to complete or for specific events to occur. & \emph{MOOSE}: ``TODO: Add a sync time; Remove after old output system is removed; sync times are handled by OutputWarehouse." & \cite{maipradit_wait_2019,vidoni_self-admitted_2021}\\
        \emph{Scientific} & Accumulation of suboptimal scientific practices, assumptions, and inaccuracies within scientific software that potentially compromise the validity, accuracy, and reliability of scientific results. & \textit{Detailed examples provided in the following sections.} & this study\\
        \bottomrule
    \end{tabular}}
\end{table*}

Throughout the categorization process, we thoroughly reviewed the documentation of the research software projects involved. Leveraging the reliability of large language models (LLMs) \cite{chiang_can_2023}, we used ChatGPT and GitHub Copilot to aid in our analysis. ChatGPT helped us understand comments with complex scientific terms and jargon by providing explanations and context for domain-specific terminology, particularly in comments related to scientific algorithms and methods. 

GitHub Copilot assisted in navigating different programming languages, understanding code structures, and identifying the intent behind code changes, especially in polyglot codebases. Importantly, LLMs were used solely as comprehension aids---to understand unfamiliar domain terminology and navigate unfamiliar code---and not to make or influence categorization decisions. All labeling judgments were made by the human coders. Additionally, for each identified SATD instance, we cross-referenced the associated documentation and user manuals to verify context and accuracy. This careful extraction process resulted in a comprehensive dataset of 28,680 labeled SATD comments, which we provide in our replication package\footnote{\url{https://github.com/AwonSomeSauce/ScientificSATD}}. %

\subsection{Findings RQ1: Cross-Domain Challenges Reflected in SATD in Research Software}
\begin{figure}[ht]
    \centering
    \includegraphics[width=0.9\textwidth]{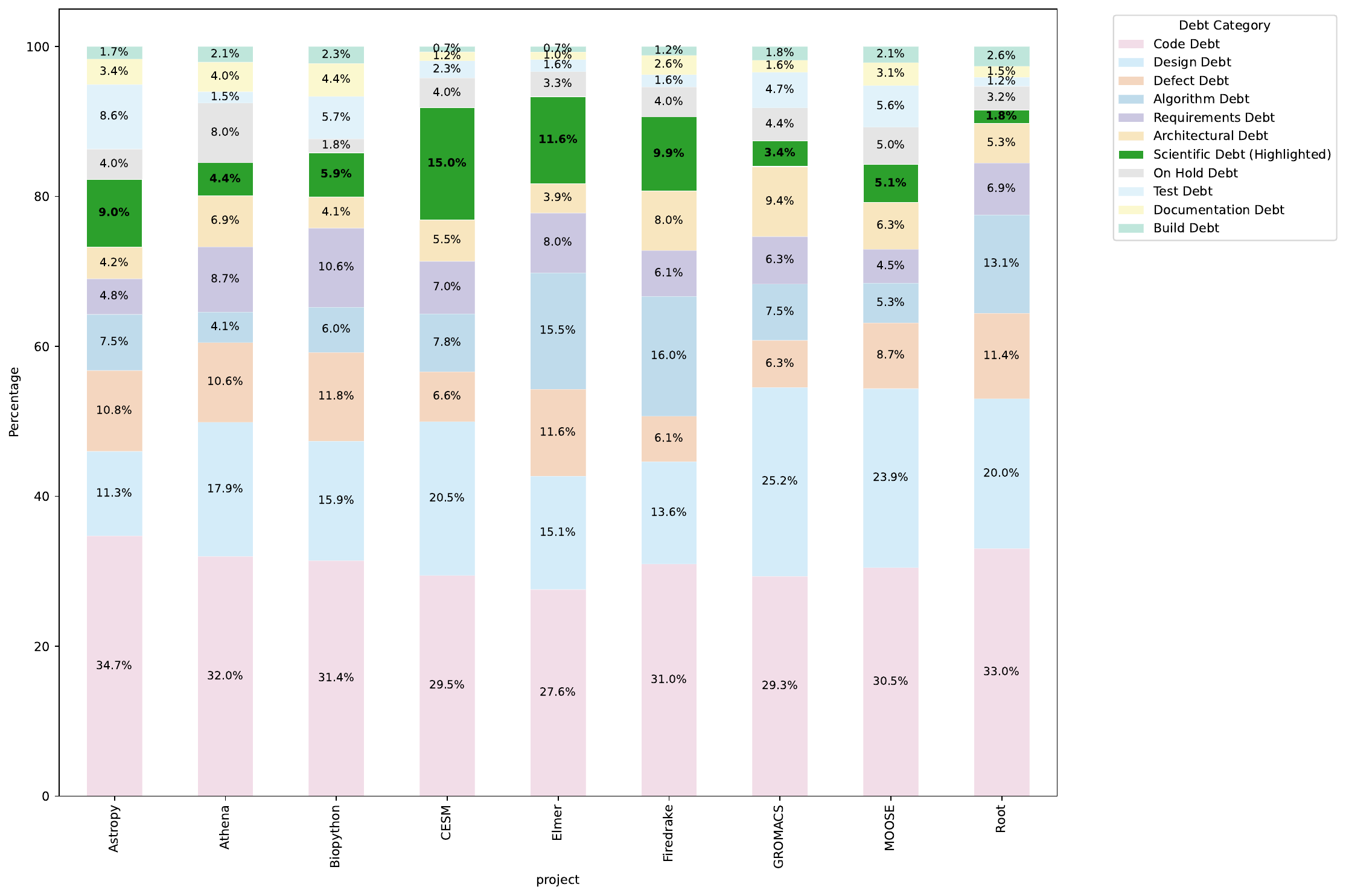}
    \caption{Percentage of SATD types across selected research projects, scientific debt highlighted.}
    \Description{Bar chart showing SATD distributions. Trends are similar, with code and design debt being most prevalent.}
    \label{fig:all_debts}
\end{figure}

To analyze the distribution of SATD categories and identify cross-domain challenges in research software, we first counted all instances of each SATD category. Comments with multiple labels were counted separately for each relevant category to accurately capture their prevalence. For instance, a comment categorized as both design debt and code debt was included in the counts for both categories. This approach allowed us to determine the representation of each SATD type within the projects. We then calculated the percentages for each category across all projects and visualized these distributions in a bar chart, as shown in Figure \ref{fig:all_debts}.

Figure \ref{fig:all_debts} illustrates the distribution of various types of technical debt across the analyzed projects, with Code Debt being the most frequently self-admitted, followed by Design Debt and Defect Debt. %
The prevalence of Code and Design Debt aligns with previous research findings on research software practices \cite{hannay_how_2009, hook_mutation_2009, kelly_analysis_2011, kelly_industrial_2013, pinto_how_2018}. This supports the reports from those studies indicating that the research software community typically prioritizes scientific objectives over robust coding and design practices. This prioritization results in compromised code quality and suboptimal design decisions, driven by the urgent need for immediate scientific results. We provide further support for this in \S \ref{sec:quaL}.

\paragraph{Scientific Debt}
\label{sec:sddefn}
A smaller portion of the self-admitted technical debt in research software projects is due to \textbf{Scientific Debt}. These  \textit{signal} underlying issues at the intersection of scientific theories and software engineering practices, complicating the integration of scientific models into computational systems. This type of debt is notably prevalent across all projects, with CESM (14.43\%), Elmer (11.16\%), and Firedrake (9.21\%) showing the highest percentages.

Unlike traditional forms of technical debt, which primarily concern software engineering issues such as code quality and design practices, \textbf{Scientific Debt} specifically arises from the challenges inherent in translating complex scientific methodologies into computational models. We define this novel category as \textit{the accumulation of suboptimal scientific practices, assumptions, and inaccuracies within research software that potentially compromise the validity, accuracy, and reliability of scientific results.} Our use of the word `science' here refers to software projects that deal with ``systematic and critical investigations aimed at acquiring the best possible understanding of the workings of nature, people, and human society~\cite{sep-pseudo-science}".

Scientific Debt highlights the unique complexities in research software development, where domain-specific intricacies and the need to align scientific knowledge with robust software engineering practices often lead to technical debt, potentially compromising the accuracy and reliability of scientific outcomes.

Scientific Debt is distinct from Algorithm Debt. Algorithm Debt, as defined by \citet{vidoni_self-admitted_2021}, concerns algorithms that are suboptimal in terms of \emph{performance or scalability}---the algorithm works but could be faster or more efficient. Scientific Debt, by contrast, concerns the \emph{scientific correctness} of the implementation: whether the software faithfully represents the underlying science. A numerically stable but scientifically incorrect assumption is Scientific Debt, not Algorithm Debt. In practice, some comments may carry both labels (e.g., an algorithm that is both inefficient and scientifically approximate), but the distinction lies in the nature of the concern: computational efficiency versus scientific fidelity.

In our labeling, we found that Scientific Debt can manifest in various forms, which we refer to as indicators:

\begin{itemize}
    \item \textbf{Translation Challenges:} Difficulties in accurately representing scientific concepts and theories within computational frameworks. This can lead to oversimplifications or incorrect implementations that do not fully capture the intricacies of the original scientific models.
    \begin{itemize}
        \item \textbf{Example from Astropy:} \textit{``We are going to share $n_{eff}$ between the neutrinos equally. In detail, this is not correct, but it is a standard assumption because properly calculating it is (a) complicated (b) depends on the details of the massive neutrinos (e.g., their weak interactions, which could be unusual if one is considering sterile neutrinos)."}
        
        In this example, the comment highlights the challenge of simplifying complex interactions between neutrinos for practical implementation in the code. The standard assumption used, though common, is acknowledged as not entirely accurate, demonstrating the trade-off between scientific precision and computational feasibility.

        \item \textbf{Example from Elmer:} \textit{``The computation of the differential of the Hencky strain function is based on its truncated series expansion. TO DO: The following involves the differential of the Hencky strain function. For some reason it doesn't appear to give convergence. Therefore, we still omit this and replace the Hencky strain differential by the differential of the Lagrangian strain. This is expected to work for reasonably small straining. Find a remedy!"}
        
        This comment points out an unresolved issue with the convergence of a specific mathematical function used in the software. The temporary solution, while working under certain conditions, highlights the ongoing struggle to achieve an accurate and reliable representation of the scientific model.
    \end{itemize}

    \item \textbf{Assumptions:} The necessity to embed assumptions within the code due to limitations in data, understanding, or computational resources. These assumptions, while necessary for initial model development, may introduce inaccuracies or biases that affect the outcomes of simulations and analyses.
    \begin{itemize}
        \item \textbf{Example from CESM:} \textit{``We assume here that new ice arrives at the surface with the same temperature as the surface. TODO: Make sure this assumption is consistent with energy conservation for coupled simulations."}
        
        This example showcases an assumption made to simplify the modeling of ice formation. The need to verify this assumption underscores the potential risk of it affecting the accuracy of energy conservation in coupled simulations, reflecting the impact of embedded scientific assumptions.
    \end{itemize}

    \item \textbf{New Scientific Findings:} The need to continually update software to reflect the latest scientific discoveries and advancements. Failure to incorporate new findings can result in outdated models that do not leverage the most current scientific knowledge, thereby diminishing the relevance and accuracy of the software.
    \begin{itemize}
        \item \textbf{Example from Astropy:} \textit{``This frame is defined as a velocity of 220 km/s in the direction of l=270, b=0. The rotation velocity is defined in: Kerr and Lynden-Bell 1986, Review of galactic constants. NOTE: should this be l=90 or 270? (WCS paper says 90)."}
        
        This comment indicates a discrepancy in the scientific constants used, with references to differing values in literature. It highlights the necessity to review and update the code to incorporate the most accurate and current scientific findings.
    \end{itemize}

    \item \textbf{Missing Edge Cases:} Limitations in the software's ability to handle all relevant scenarios or edge cases. This can lead to incomplete or erroneous results, particularly in complex scientific domains where edge cases may have significant implications.
    \begin{itemize}
        \item \textbf{Example from ROOT:} \textit{``This does not work for large molecules that span > half of the box!"}
        
        The comment points out a limitation in the software's capability to handle large molecules, which could lead to significant inaccuracies in simulations involving such cases. It underscores the importance of ensuring comprehensive coverage of all possible scenarios to maintain the reliability of the software.
    \end{itemize}

    \item \textbf{Computational Accuracy:} Instances where the mathematical or scientific accuracy within the software is compromised. This can occur due to simplifications, numerical precision issues, or incorrect implementation of scientific algorithms, leading to unreliable or incorrect results.
    \begin{itemize}
        \item \textbf{Example from GROMACS:} \textit{``TODO: For large systems, a float may not have enough precision."}
        
        This comment highlights a concern with numerical precision in large systems. The use of float data types may lead to significant inaccuracies, indicating a need for better precision management in scientific computations.

        \item \textbf{Example from GROMACS:} \textit{``Since the energy and not forces are interpolated, the net force might not be exactly zero. This can be solved by also interpolating F, but that comes at a cost. A better hack is to remove the net force every step, but that must be done at a higher level since this routine doesn't see all atoms if running in parallel. Don't know how important it is? EL 990726."}
        
        This comment describes a potential issue with force calculations due to interpolation methods used. The proposed hack to address the issue indicates a temporary workaround, highlighting the compromise in scientific accuracy and the need for a more robust solution.
    \end{itemize}
\end{itemize}

\begin{figure}[ht]
    \centering
    \includegraphics[width=.9\textwidth]{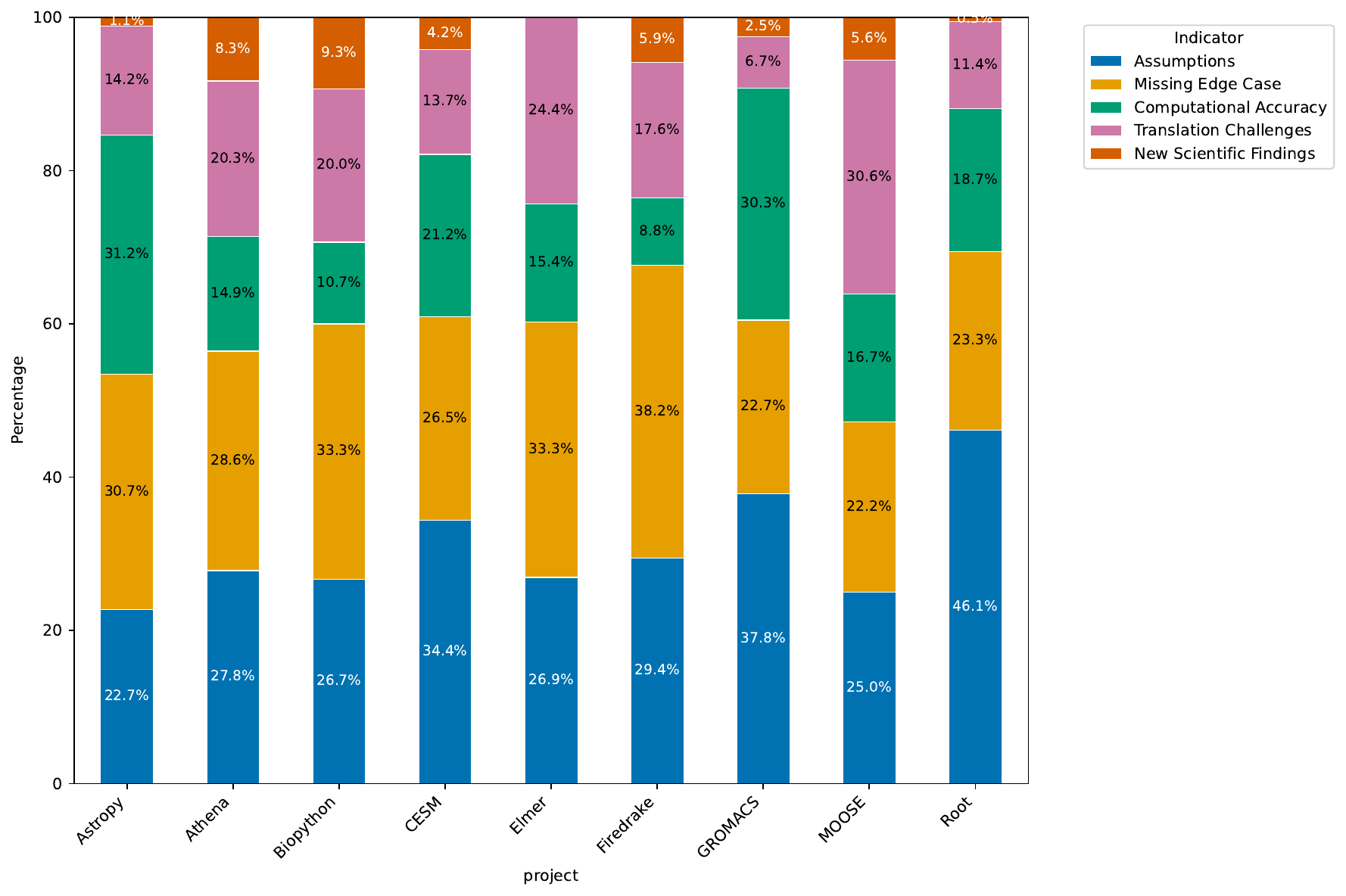}
    \caption{Percentage of Scientific Debt indicators across research software}
    \Description{Stacked bar chart showing indicator prevalence.}
    \label{fig:all_scientifics}
\end{figure}

Figure \ref{fig:all_scientifics} shows the distribution of Scientific Debt indicators across the analyzed projects, with Assumptions and Missing Edge Cases being the most frequent. For example, GROMACS has 37.82\% assumptions, and CESM has 34.53\%. These high percentages suggest that these projects often rely on assumptions to simplify complex phenomena or compensate for limited data and computational resources, reflecting a pragmatic approach to advancing scientific inquiry. 

Similarly, missing edge cases are prevalent, particularly in Firedrake (38.24\%) and Elmer (33.33\%). This indicates significant challenges in handling all possible scenarios within these projects, highlighting the difficulties in comprehensively testing and validating research software, which often deals with highly variable data and complex phenomena.

Computational accuracy issues are particularly prominent in projects like Astropy (31.25\%) and GROMACS (30.25\%), underscoring the continuous challenge of maintaining numerical precision and reliability in scientific computations.
Translation challenges are significant in projects such as MOOSE (30.56\%) and Athena (20.33\%), highlighting the complexity of converting theoretical scientific models into practical computational algorithms, which reflects the intricate nature of research software development.

\begin{tcolorbox}[colback=yellow!10!white, colframe=yellow!40!black, title=Summary]
    While Code Debt and Design Debt are the most prevalent types of SATD in research software, we identified a novel category called Scientific Debt. This category, significant across all projects, includes indicators such as assumptions, missing edge cases, computational accuracy, translation challenges, and new scientific findings.
\end{tcolorbox}

\section{Study 2: Perceptions of Technical Debt}
\label{sec:quaL}
We now look to explain the results from the results in Study 1 using our next method, an interview study of research software developers. This second phase builds on the insights we gain into research software and the occurrence of a new form of self-admitted technical debt, Scientific Debt, to answer our second research question:
\noindent\textbf{RQ2: How do practitioners in research software projects perceive Technical Debt?}

\subsection{Research Method}
\subsubsection{Participant Selection and Interview Process}
We used personal connections and purposive sampling from cold calling \cite{baltes_sampling_2022} to create an initial sample of participants, shown in Table \ref{tbl:interviewees}. Our sample was drawn from a selection of projects with some overlap with projects in the RQ1 portion of the paper. Five participants contribute to projects from RQ1.

Cold calling has the drawback of not being representative of all researchers involved in research software, but the advantage of being tailored to people with meaningful involvement in the project, as well as mapped to the project domains covered in RQ1. The exploratory nature of the interviews and our constructivist philosophy align well with these tradeoffs.

Note that due to the potential for identification, and per our ethics board, we have not included project names but only the science domain.%

\begin{table}[bh]
    \caption{Interview Participants. * - most common language.}
\begin{tabular}{llllll}
\toprule
\textbf{Ref} & \textbf{Domain} & \textbf{Role} & \textbf{Language*} & \textbf{Context} & \textbf{Training} \\
\midrule
P1 & Climate & Developer & Fortran & prod & CS \\
P2 & Climate & Developer & Fortran & prod & CS \\
P3 & Climate & PostDoc & Fortran/R & explore & Domain \\
P4 & Climate & Developer & Fortran & prod & CS \\
P5 & Climate & PI & Fortran & explore & Domain \\
P6 & Astronomy & Maintainer & Python & prod & Domain \\
P7 & Molecular Dynamics & Developer & Python/C++ & prod & Domain \\
P8 & Climate & PI & Fortran/C++ & prod & Domain \\
P9 & Physics & Maintainer & C++ & prod & Domain \\
P10 & Mathematics & Maintainer & C++ & prod/explore & Domain \\
P11 & Climate & Developer & Fortran/C++ & prod & Domain \\
\bottomrule
\end{tabular}
\label{tbl:interviewees}
\end{table}

 The interview followed a semi-structured approach using the questions outlined in Table \ref{tbl:questions}. 
Although we designed an interview guide, its purpose was to guide the conversation
while allowing fluidity by prompting the interviewee to elaborate and explain further based on their responses.
As interviews proceeded, we added some questions (bottom of table) to cover emerging themes of interest. In particular, the questions about specific examples of technical debt (``What caused this specific example?'', ``What were the consequences?'') and the section on incorporating domain knowledge and scientific models were designed to elicit the kinds of cross-domain challenges identified as Scientific Debt in Study 1.

\begin{table}[]
   \caption{Key Parts of the Interview Guide}
\begin{tabular}{p{\textwidth}}
\toprule
\textbf{Experiences and personal history} \\ \midrule
   What does the project do? \\
   What are its goals? \\ 
Amount of time dedicated to programming vs science.? \\ 
What mentorship or learning is provided for newcomers? \\ \midrule
\textbf{Technical debt} \\ \midrule
Looking at the software you maintain, but also other code you use, what would you say are the main challenges?\\
   Have you heard of technical debt? If yes, can you give me your definition? \\
   Do you see examples of technical debt in the code you work on? \\
   What caused this specific example? \\
   What were the consequences of this example?  \\
   How do you remediate problems with your code? \\
   What tools, if any, do you find helpful?  \\ \midrule
\textbf{Incorporating domain knowledge and scientific models} \\ \midrule
   Scientific software often has a heavy domain component (e.g., climate modeling). \\
   
   How well is that integrated with your code?  \\
   Do you take any particular steps to manage this? \\ \midrule

\textbf{Added questions after P6}\\ \midrule
How does your project use Github or other tools for maintenance? 
How has the design changed/would it be changed? 
What is that process like? \\

\bottomrule
\end{tabular}
\label{tbl:questions}
\end{table}

\subsubsection{Data Analysis and Coding}
All interviews were conducted using Zoom and lasted 40-75 minutes. 
The first author conducted the interviews over the months from October 2023 to June 2024.
We used Otter.AI to transcribe the interviews, and the first author then went through the transcripts to correct the AI's errors and to anonymize the transcripts, which we release in the replication package. 

We adopted a constructivist stance~\cite{Easterbrook2008SelectingEM}, which aligns with our objective to understand research software engineers and their perceptions and experiences with technical debt.
By adopting this stance, we recognize that knowledge and meaning are actively constructed by individuals through their experiences and interactions.

Technical debt is a well-established construct in the research literature, but the nature of technical debt in the context of research software is not. We thus conducted a combined deductive and inductive coding exercise. We began with a deductive codebook derived from the literature (Section~\ref{sec:deductive}); during coding, when a passage did not fit an existing deductive code, we created a new \emph{de novo} code to capture the idea. Both deductive and inductive codes were applied simultaneously in a single pass through each transcript. The four themes reported below emerged from collating all codes---both deductive and inductive---into higher-level groupings during the thematic analysis.

\begin{table}[h]
\centering
\caption{Sample quotes, codes, and themes showing traceability to original data. Complete examples available in replication package.}
\begin{tabular}{p{6cm}p{1cm}p{3cm}p{2cm}}
\toprule
Quote & \textsf{Code} & Definition & Theme \\
\midrule
"I had the same program run on 2 different workstations, both Silicon graphics machines. Just one year difference. And I get different numbers out. And there were 5\% differences. And the method was supposed to be exact to 0.5\%. & \textsf{interest-science} &	scientific accuracy suffers as a result of TD & Science \& Org. Goals \\
"I brought some piece of Fortran code and I sent together with my office neighbors who was working [on] slightly different program writing his own Fortran code, and we exchanged and checked each other" & \textsf{process-code-review} & team does code reviews & Boundary Obj.\\
``Is this like a critical change that we really need for this upcoming simulation? Or is this ... would be nice to have ... and where does this fall on the priority list? & \textsf{cause-business-priority}	& priority of features over maintenance - refactoring less important   & Science \& Org. Goals \\
``[where] were the emissions were happening within the code, but once you find that, find out where it's computing the flux" & \textsf{context-code-structure}	& Statements about how the code is architected/designed that are not quality judgments & Complexity\\
``All my Git knowledge is very limited. I just learned it on the on the go" & \textsf{rse-training} &	the RSE is doing the work or started the work as part of education & Team \& People  \\
\bottomrule
\end{tabular}
\label{tbl:interview_codes}
\end{table}

Table \ref{tbl:interview_codes} highlights a few examples of each type, with the complete code book available as a supplement. For example, the notion of \emph{technical debt interest}, that is, the ongoing payments caused by technical debt, was introduced in \citet{kruchten_technical_2012}. As we coded the interviews, however, patterns emerged that were not covered by the existing codes. These \emph{de novo} codes are flagged as such in our codebook. 

\noindent\textbf{Coding.} We followed the thematic analysis guidelines of \citet{braun_using_2006}. That method consists of 
\begin{enumerate}
    \item Familiarization: re-reading transcripts and audio to deeply engage with the data. Some initial ideas noted.
    \item Initial Coding: Systematic low-level coding across all transcripts, expanding the code book as necessary.
    \item Searching for Themes: Collate codes into potential themes, alongside relevant data.
    \item Theme Review: ``Checking if the themes work in relation to the coded extracts (Level 1) and the entire data set (Level 2), generating a thematic `map' of the analysis."~\cite[p. 87]{braun_using_2006}
    \item Define and Name Themes: Refine specifics of each theme and overall narrative, ensuring clear theme labels.
    \item Report Production: Extract representative quotes and relate back to the relevant literature.
\end{enumerate}

We used the open source tool Taguette~\cite{Rampin2021} to organize the transcripts and coding for the data. We provide Taguette's exported SQLite DB in our replication package. This database contains the transcripts and our codes. 

Upon the completion of the analysis, we organized member checking ~\cite{milesQualitativeDataAnalysis2014} to collect feedback from our interviewees on our findings, reported in Sec \ref{sec:member}.

\subsection{Findings RQ2}
We report findings in two ways. For our deductive coding exercise, based on pre-existing theories about technical debt, we present a simple descriptive summary of code frequency, alongside those codes which were theorized but not seen. Then we present a more detailed, inductive assessment of our interviews by showing a thematic analysis of the data, grounded in the source material. 

\subsubsection{Deductive Coding Results}
\label{sec:deductive}
We applied deductive coding using predefined theory constructs from the papers of \citet{pinto_how_2018,ernst_technical_2021,alves_identification_2016,rios_most_2018,martini_architecture_2014}. Our replication package contains a table mapping all of our codes to this existing work for traceability. 

Table \ref{tab:codes_ded} reports on the codes we deductively coded that had more than 10 occurrences. 

\begin{table}[h]
\centering
\caption{Codes from literature with 10 or more occurrences. Part.: number of distinct participants (out of 11) who contributed the code. RSE: research software engineer.}
\begin{tabular}{p{2cm}p{5cm}ccl}
\toprule
Tag & Description & Freq. & Part. & Source \\
\midrule
cause-business-priority & Priority of features over maintenance & 32 & 11 & \cite{martini_architecture_2014} \\
cause-social-silos & Silos in org & 24 & 8 & \cite{ernst_technical_2021} \\
cause-social & Social factors cause TD & 23 & 9 & \cite{martini_architecture_2014} \\
cause-techevo & Reliance on legacy hardware or software & 23 & 5 & \cite{martini_architecture_2014} \\
rse-pain-reward & Lack of formal reward (publications) & 23 & 6 & \cite{pinto_how_2018} \\
rse-pain-lack of time & RSE has no time & 21 & 8 & \cite{pinto_how_2018} \\
interest-rework & Rework has to be done due to TD & 20 & 8 & \cite{rios_most_2018} \\
rse-pain-mismatch & Mismatch between coding and domain skills & 19 & 9 & \cite{pinto_how_2018} \\
cause-business-time & Time pressure & 17 & 7 & \cite{martini_architecture_2014} \\
cause-docs & Incomplete or poor docs & 17 & 5 & \cite{martini_architecture_2014} \\
cause-design-arch & Shortcuts in design and arch choices from before cause TD & 16 & 7 & \cite{ernst_technical_2021} \\
cause-lack of knowledge & Didn't know enough (that we should have known) & 16 & 6 & \cite{rios_most_2018} \\
interest-indicator-doc issues & TD is causing documentation problems & 15 & 6 & \cite{alves_identification_2016} \\
cause-req & Ignorance of requirements, poor requirement, no elicitation done, bad user stories & 14 & 7 & \cite{martini_architecture_2014} \\
interest-low quality & Software has low quality & 12 & 8 & \cite{rios_most_2018} \\
rse-pain-collaborate & Hard to collaborate on software project & 12 & 6 & \cite{pinto_how_2018} \\
cause-test & Poor or incomplete tests & 12 & 4 & \cite{ernst_technical_2021} \\
interest-low maintainability & Software hard to maintain & 11 & 5 & \cite{rios_most_2018} \\
rse-pain-docs & Poor docs bother RSE & 10 & 6 & \cite{pinto_how_2018} \\
\bottomrule
\end{tabular}
\label{tab:codes_ded}
\end{table}

Additionally, the below codes which were pre-existing, i.e., described in other research publications, had fewer than two occurrences. %
\begin{itemize}
    \item \textsf{rse-pain-user}~\cite{pinto_how_2018}: the pain from dealing with users. This might reflect that users may not be as relevant if you are the user.
    \item \textsf{rse-pain-interruptions}~\cite{pinto_how_2018}: we did not see examples where interruptions caused pain, perhaps reflecting a less user-driven process.
    \item \textsf{interest-financial loss}~\cite{rios_most_2018}: interest payments on debt may be causing financial loss. We did not observe this, but it is perhaps less relevant when the focus is the science.
    \item \textsf{interest-indicator-static}~\cite{alves_identification_2016}: static analysis issues as another source of interest costs; in our respondents, few use static analysis tools.
    \item \textsf{cause-incompleteref}~\cite{martini_architecture_2014}: incomplete refactoring as a cause of TD. We found little evidence of refactoring.
    \item \textsf{cause-reuse}~\cite{martini_architecture_2014}: reuse as a cause of TD. While this theme did not occur, we posit it is so common it was not even mentioned in our interviews.
\end{itemize}

\subsubsection{Inductive Analysis}
\begin{figure}[ht]
    \centering
    \begin{subfigure}[b]{0.9\textwidth}
        \includegraphics[width=\textwidth]{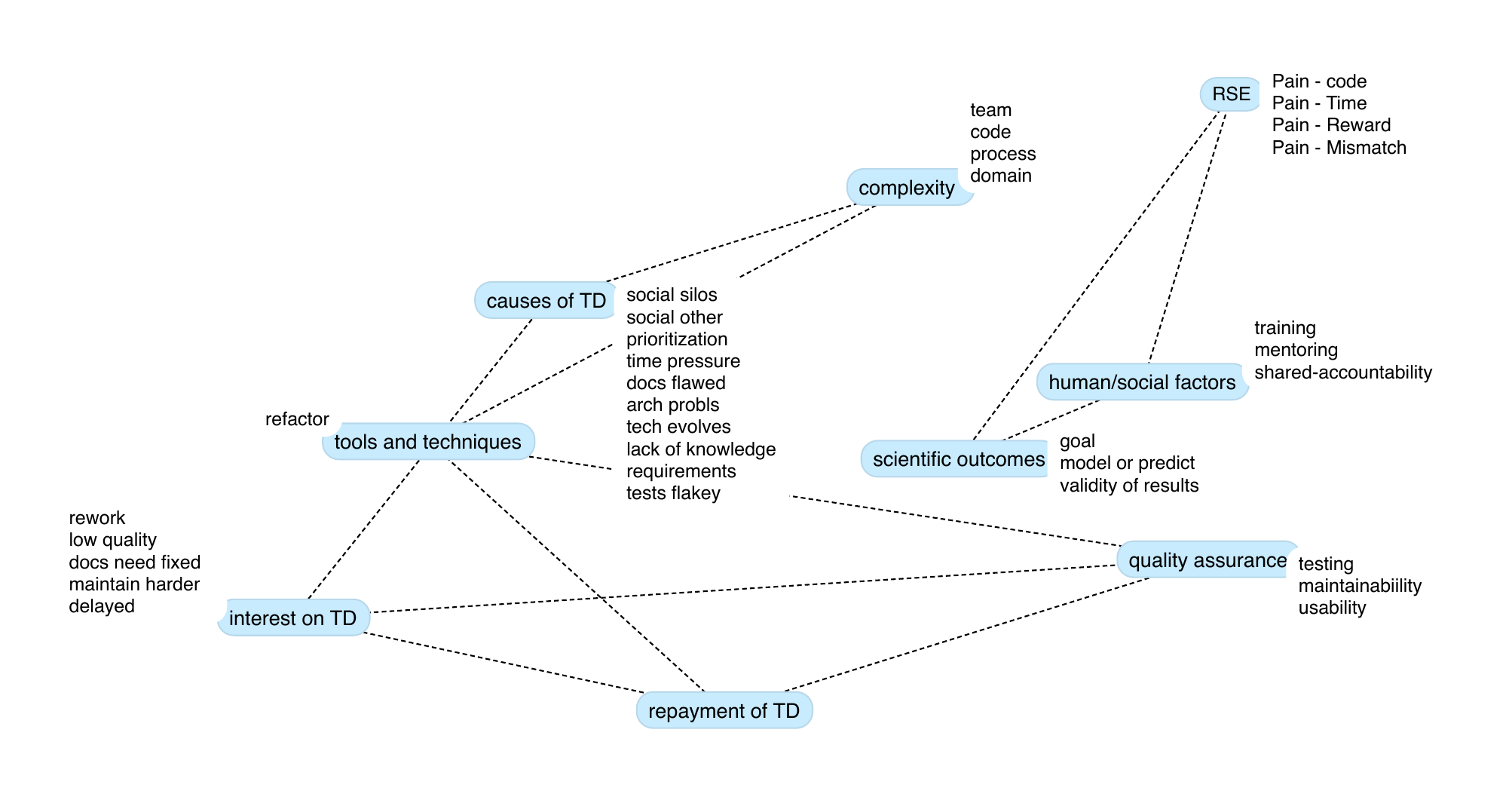}
        \caption{Initial thematic map.}
        \label{fig:mapv1}
    \end{subfigure}
   
    \bigskip

    \begin{subfigure}[b]{0.75\textwidth}
        \includegraphics[width=\textwidth]{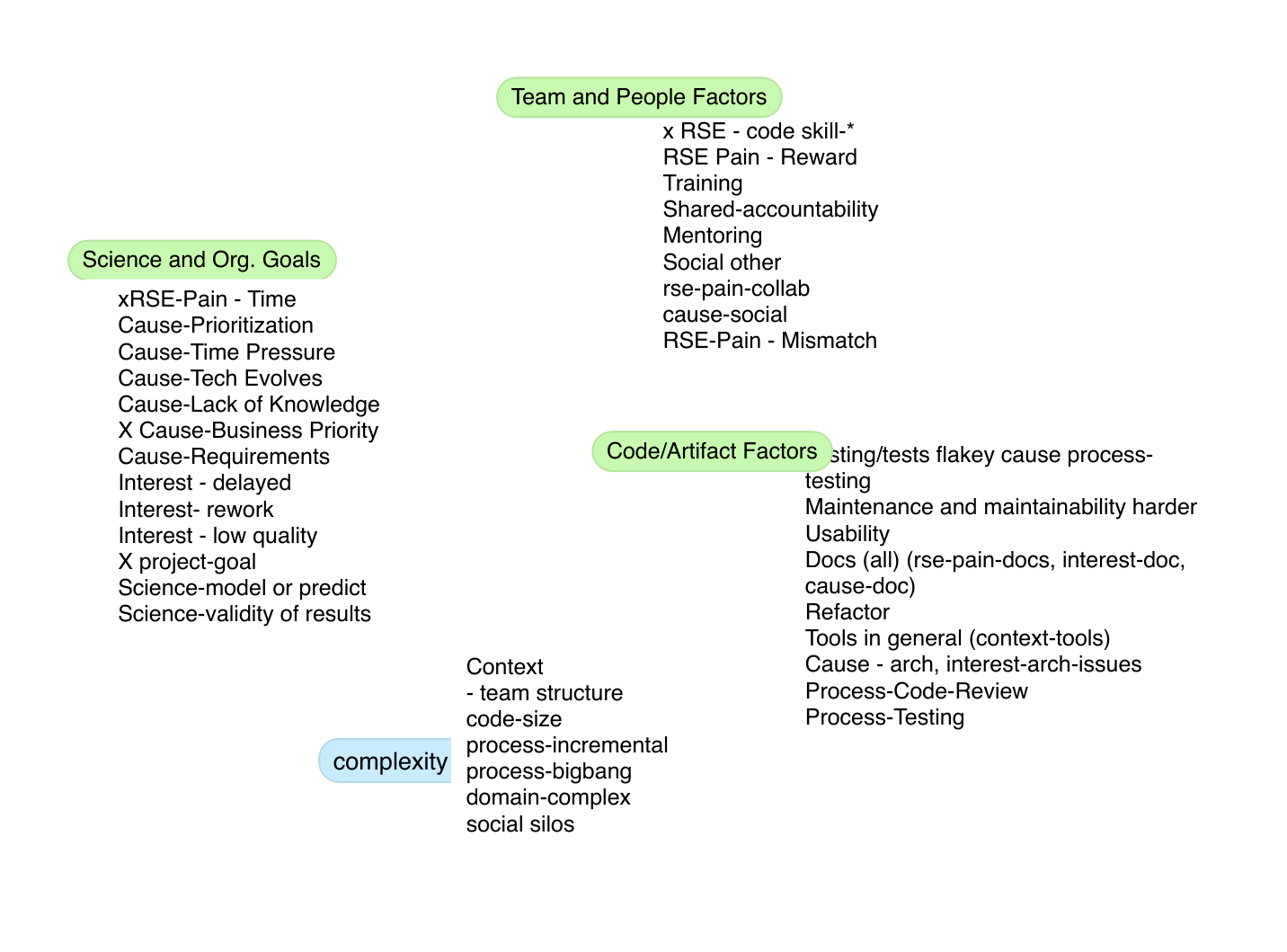}
        \caption{Revised thematic map, after coalescing around key themes.}
        \label{fig:mapv2}
    \end{subfigure}
    \caption{Thematic maps showing how our analysis evolved.}
    \Description{Two thematic maps. The first (initial) shows codes loosely grouped around causes of TD, interest on TD, and human/social factors. The second (revised) consolidates these into four themes: Artifacts As Boundary Objects, Science/Organization Goals Drive TD Management, People As Drivers of Success, and Complexity Complicates.}
    \label{fig:thematic-map}
\end{figure}

We identified four overarching themes from our inductive analysis. These are \textsf{Artifacts As Boundary Objects}, \textsf{Science/Organization Goals Drive TD Management}, \textsf{People As Drivers of Success (In Spite of Barriers)}, and finally, \textsf{Complexity Complicates}. We elaborate on these with reference to specific quotes and low-level codes below. Figure \ref{fig:thematic-map} highlights how our codes and themes evolved. In the initial map (\ref{fig:mapv1}), codes clustered around broad groupings such as causes of TD, interest on TD, and human/social factors. As we iterated, we consolidated these into four themes: for example, codes related to scientific outcomes, organizational priorities, and time pressure coalesced into \textsf{Science/Organization Goals Drive TD Management}, while codes about team structure, domain complexity, and code size merged into \textsf{Complexity Complicates}. Several initial groupings (e.g., ``repayment of TD,'' ``tools and techniques'') were absorbed into other themes rather than standing alone.

\noindent\textbf{Theme 1: Artifacts As Boundary Objects} Artifacts are boundary objects that facilitate collaboration. Boundary object theory~\cite{starInstitutionalEcologyTranslations1989,wohlrabBoundaryObjectsTheir2019a} posits that inter-team communication often revolves around artifacts, ``a means of enabling collaboration between different groups of actors''. These boundary objects ``inhabit several intersecting social worlds and satisfy the informational requirements of each of them"~\cite{starInstitutionalEcologyTranslations1989}.  In addition to formal communication artifacts such as emails and meeting notes, boundary objects are artifacts which these different groups shape together, often in what \citet{starInstitutionalEcologyTranslations1989} called \emph{standardized forms}. In our study, these objects include source code, code management items such as pull requests, specifications, standards, and design documents. Each group understands the artifacts, but specializes it for their specific purposes.

A boundary object view is important in these teams, which are inherently cross-disciplinary according to the  domains of~\citet{kelly_scientific_2015}. For example, ``dealing with scientists and software engineers who both think they're talking about the same thing, but haven't troubled to find out whether they are and wondering why they're talking past each other (P8 27 minutes)". Artifacts make this communication easier, as there is \textbf{a single source of truth} (e.g., the code). 

Across projects, the artifact is important because it is tangible and what people do or talk about. ``I can hand somebody a PDF and say, This is the spec for that data file format (P6 57m)".
For a lot of interviewees, the \textbf{code is the main boundary object}, and code review is key to forming understanding across teams: "instead of writing the formulas and Latex, it's actually faster for me to write them in Python, and then have it as a pull request, as opposed to have somebody else type it off again (P6 57m)". Code artifacts act as a visible manifestation of technical debt, both pain and cause: "we've got to find the right scientists to talk to and we'll sit down together and see what we can to puzzle out about how this is working. (P6 46m)"

The overall architecture is a key interaction shaper: "we have split out things out of the core and separate into separate packages to be installed, separate and to be updated on other [teams's] schedules. (P6 62m)". Architecture is an artifact directly influencing the way science is done, e.g. with grid models: "we get to a point where we can do changes to small pieces without accidentally breaking other parts of it because of the more isolated and more you have cleaner interfaces. (P6)"

One unique artifact was the scientific paper, with venues to accept publications focusing specifically on the software driving the discoveries (e.g., \emph{Journal of Open Source Software}.): "I should actually write that paper, I mean, [I have] given talks about it (P8)". Different teams across organizations then read these experience reports to learn more about design choices.

Boundary objects likely play a dual role with respect to technical debt. They make any debt visible and actionable (e.g., through code review of Scientific Debt comments, such as those described in \S \ref{sec:sddefn}). 
However, when these artifacts are poorly maintained or misunderstood, they can also become sources of debt themselves. In several interviews participants referenced code that was essentially unmaintainable since the author had left the project: ``we lost our main font of knowledge about how we reconstructed electrons (P9, 25m)''.

\vspace{.2cm}
\noindent\textbf{Theme 2: Science/Organization Goals Drive TD Management}
We found that most of the challenges with managing technical debt in research software are caused by a \textbf{combination of science and organizational goal setting}. These goals often conflict with the need to manage technical debt and make the software maintainable.

We did not see a lot of evidence in the interviews that individual actions are the problem, such as bad coding practices.
Instead, changing priorities influence the code/artifacts: "there's a lot which we want to do, which is not done because the highest priority and the really critical stuff is what gets done (P9 43m)". 

Science projects are discovery oriented, and \textbf{work that does not support new discoveries is not a priority} for the leadership or funders.
"National Science Foundation grants, you know, they target some particular development. And then they just expect that, Okay, that's done. (P1 16m)". Most projects are lead by domain experts who may not know much about software: "we had a period of time when a significant fraction of the leadership couldn't write a single line of Python code" (P6 6m); the "model has a scientific steering committee that, you know, basically makes decisions about what what's going to be worked on and that percolates down. (P1 23m)". This mimics financial pressures driving technical debt in commercial settings~\cite{ernst_measure_2015}.

Projects have goals and mandates \textbf{from within and from without/above}: "the physics performance is ultimately what people care about (P9 42m)". However, these goals typically do not include long-term maintenance, despite the team understanding this to be a significant challenge. P2: "lots of technical debt and that we want to come back and clean up and and so staying on top of those issues is its own challenge (P2 49m)." "Often, I think in scientific, like code development, we do not take into the account the overhead of optimum optimizing that code or like, improving like the software engineering aspect of that code (P11 10m)." . As a result of misunderstanding software, "[leadership] don't see that software is a dynamic artifact that it changes over time, and the results change over time. Even the methods implemented change over time (P7)".

The project \textbf{cadence} is also science driven, e.g., to align with worldwide collaborations such as the Climate Model Intercomparison Project (CMIP) or start up of costly major instruments, like the CERN Large Hadron Collider: "It's been a bit close really for data taking this year, especially as the trigger is complicated and we use it in a very advanced way (P9 13m)." 

The lack of software awareness means projects are often reactive: "So we can't really leave it because because they'll be deploying them with the latest Pytorch or whatever, alongside it. And so if we get too far away from the bleeding edge, we'd be in trouble (P8 12m)". 
As a result, projects face interest payments in the form of \textbf{delayed releases or difficult to change code}. "I think we know when it's [tech debt] happening, because just the, there's more friction when you work on a piece of software and kind of run into things, problems (P4 21m)", and "in the process of adding stuff we are forced essentially to, to address some of the underlying technical debt problems, because otherwise we can't implement the new stuff because it just won't work (P6 35m)."

The project then needs to spend time reworking the code: "there's at least a dozen spectroscopy reduction pipelines that fundamentally do the same thing, but implemented by different people for different instruments (P6 45m)", which impacts science goals: "we kind of got left with this thing that was thrown together really hastily, but everyone was really anticipating the output from it (P4 16m)". Ideally, the interest in the form of bugs are at least visible when they impact science goals: "Our best bugs are the ones that the model just blows up and crashes. And it's really obvious. But you know, what we really work hard to avoid, are the bugs that like, the scary bugs, or the bugs where everything runs fine. It looks like it's working, right. But it's the things are working wrong (P2 61m)".

Notably, while our Study 1 findings show that Scientific Debt is present across all projects, our interviews reveal that it is rarely prioritized for remediation. Scientific Debt is paradoxically driven by the same science goals that deprioritize its removal: the pressure to produce scientific results creates the debt, while the pressure to produce the \emph{next} result defers its repayment.

\noindent\textbf{Theme 3: People As Drivers of Success (In Spite of Barriers)}
In spite of the challenges to project success from Theme 2, our interviews found that the individuals developing research software \textbf{exhibit high technical and domain skills}. 
Individuals are the most dominant signal in project success: "we've been fortunate to have staff that are here for longer term and take a long term view and are able to design things to be sustainable (P4 15m)" and have deep knowledge of the code: "I've seen every sub module I've seen every every major import (P6 65m)". For science-dominant careers, coding is not always complicated, if the project lives a long time: "there's a module in the community atmosphere model, that fluxes sea salt intake atmosphere... And so we just add a term (P3 16m)".

There are some skill gaps, around user experience design: "we kind of just don't do user experience design (P4 13m)", and high performance specific code parallelization: "I have been interested at some point in like, parallel programming, like using MPI and stuff like that, because I didn't feel like I know much about it (P5)" or platform specific optimizations (e.g., for GPU programming).

A big challenge is turnover and recruitment, because \textbf{technical debt only matters in the context of the social system} building it (skills, structures, incentives): "a couple of super experts move on to different fields in the last couple of years. And also [in 2015], I can think of at least a few people who left who took with them huge amounts of domain knowledge (P9 25m)".

Collaboration is essential since "no one is anyone's direct boss (P9 20m)" and "we've gone past the point where any one group can own a climate model and all the components and we have to depend on each other, a lot more(P8 45m)", and a career in research software is now less of a penalty: "they allowed people to move between science track and software engineering track without being penalized in terms of career terms (P8 39m)".

The people on these projects typically are \textbf{trained on the job}: "the biggest things that I probably had to pick up were just how to work with a large code base, and how to understand a large code base (P2 21m)", which pose challenges for managing the essential complexity we document in Theme 4.

\noindent\textbf{Theme 4: Complexity Complicates}.
This theme captures the technical debt issues in research software that arise due to complexity of various kinds (such as process, domain, the code itself). 
Some of the technical debt that happens is unavoidable, caused by what Brooks called ``essential complexity"~\cite{fred_p_brooks_no_1986}.
Essential complexity varies by domain, and reflects the challenges of working at advanced levels in modern science, be it in physics, math, biology, to name a few.
It also reflects the complexity of the underlying legacy code, often decades old. 

This theme also captures what Brooks called \emph{accidental complexity}, `own goals' if you will, for example, around how teams are structured, or the portability of super computers. None of that complexity is inherent, but tends to emerge over time, like biofouling, the sea life and barnacles agglomerating onto the bottom of a ship. 

We found essential complexity existed primarily when the
\textbf{domain is complex}: "I don't think documentation alone is going to save us because this is a really very complicated domain specific problem, which we need people to learn over many years, I don't think there's any any other way of doing it other than having people operating the [high energy physics] trigger. (P9 31m)"

Respondents also identified complexity that arose when \textbf{code is enormous}: "These are like, really big codes, like more than million lines of code and, like, not easy to maintain, longer term. (P11 25m)"; for new contributors, "There's a huge amount of technical overhead (P9 7m)". Some of this complexity can be reduced with modern tools: "never been easier in that sense also for fewer people to maintain the code base, given that the tooling (P9 23m)", and therefore, accidental, but complex domains seem to require larger and more complex codebases.

Other complexity issues came from organizational factors, such as how they \textbf{fund the project} "So that's part of why, in addition to internal change, the whole funding is shifted between groups (P6 58m)" or the \textbf{process of collaborating} on the software: "we operate as a collaboration of 5000, technically independent scientists (P9 52m)".

From the operational knowledge domain, a lot of complexity exists at the hardware/software interface for supercomputing, as \textbf{portability is complex}: "you know, the problems are more Cray, and Intel's compilers are as flaky as hell. MPI is flaky on new platforms we're always dealing with with the the next version, the compiler won't work out, why why not? (P8 12m)".

Another complexity was \textbf{testing}: "test passing is relative to some baseline. And so you need to be careful about how you define that baseline and ... whether we expect that large suite of tests to pass and which ones we expect to pass (P2 32m)"

Finally, \textbf{complexity impacts science and shared understanding} "so if we rewrite the whole thing, it will be easier than figuring out like, what's going on here? (P11 37m), reflecting complexity from legacy code and poor documentation.

\section{Discussion}
\label{sec:discuss}
In our quantitative study (RQ1), we analyzed 28,680 SATD comments drawn from nine long-lived research software projects spanning astronomy, climate modeling, molecular dynamics, high-energy physics, and applied mathematics. Beyond the expected prevalence of Code Debt and Design Debt, we identified and defined a novel category, \emph{Scientific Debt}: the accumulation of suboptimal scientific practices, assumptions, and inaccuracies that can compromise the validity and reliability of scientific results. Scientific Debt manifests as unvalidated assumptions, missing edge cases, computational accuracy trade-offs, translation challenges between theory and implementation, and outdated scientific knowledge embedded in code.

In our qualitative study (RQ2), interviews with 11 contributors to long-lived research software projects revealed four themes characterizing how technical debt arises and persists. Artifacts such as source code and pull requests serve as boundary objects bridging domain scientists and software engineers. Science and organizational goal-setting, driven by publication pressure and grant cycles, routinely subordinate long-term maintainability to immediate scientific results. Despite these structural barriers, the high technical and domain skill of individual contributors is the primary driver of project success, though turnover creates fragility. Finally, both essential complexity inherent in advanced science and accidental complexity from team structure, legacy code, and platform portability continuously compound the debt burden.

The rationale for selecting a mixed methods research (MMR) design is to be able to conduct second-order inferences from both the quantitative and qualitative studies. Although we reported our \emph{findings} in each of Sections \ref{sec:quaN} and \ref{sec:quaL}, we now report a \emph{results-based integration}~\cite{storeyGuidelinesUsingMixed2024} of these findings into second order inferences about technical debt in research software. We identify three key aspects: the human and socio-technical nature of research software; managing scientific technical debt; and dealing with research artifacts in technical debt efforts. Neither study alone could produce these inferences: the quantitative study revealed \emph{what} types of debt exist and their relative prevalence, but not why they persist or how practitioners experience them; the qualitative study surfaced the organizational, social, and complexity-driven forces behind technical debt, but could not quantify its extent or distribution across projects. Together, they show that Scientific Debt is not merely an artifact of comment-mining---it reflects real tensions between scientific goals and software quality that practitioners actively navigate.

\subsection{Knowledge Domains in Domain Knowledge-Intensive Software}

\begin{figure}[ht]
    \centering
    \includegraphics[width=0.5\textwidth]{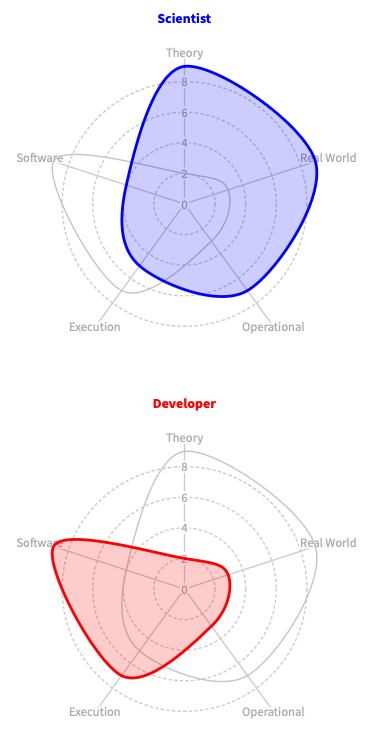}
    \caption{Illustrative knowledge profiles for two hypothesized team member archetypes, based on Kelly's domains~\cite{kelly_scientific_2015}. These are not empirical measurements.}
    \Description{Spider (radar) diagram with axes for Theory, Real-World, Software, Execution, and Operations. The ``Scientist'' profile scores high on Theory and Real-World; the ``Developer'' profile scores high on Software and Execution.}
    \label{fig:spider}
\end{figure}

Our findings on Scientific Debt (\S \ref{secch3:categorizeSATD}) in research software complement the Kelly knowledge acquisition model \cite{kelly_scientific_2015}. Kelly's model emphasizes continuous knowledge acquisition and integration, highlighting the need for deep domain-specific knowledge and systematic approaches in research software  development. 

Fig. \ref{fig:spider} illustrates a hypothetical application of the Kelly domains to two hypothesized team members. ``Scientist'' would represent an individual holding an advanced degree in a science domain, who learns software development on the job. This is most research software developers~\cite{hannay_how_2009}. ``Developer'' represents team members with training in software engineering, e.g., through a CS degree. The spider diagram captures the amount of knowledge and skill on the different domains. Developer is more skilled at Software and Execution of that software; Scientist is skilled in the science (Theory) and Operation of the software (e.g., in making climate predictions). While the diagram suggests these are two individuals, our interviews in RQ2 found that these roles can be fluid, with people moving between domain and software roles periodically. 

Both our study and Kelly's model emphasize the importance of domain-specific knowledge in research software development. We identified a novel category of technical debt, termed \textit{Scientific Debt}, with indicators like assumptions, missing edge cases, computational accuracy, translation challenges, and new scientific findings. This aligns with Kelly's emphasis on Real-World and Theory-Based Knowledge, highlighting the necessity for developers to deeply understand the scientific problems and principles behind the software. In RQ2, we found that it is the human developers that drive the project along, and the loss of any one can be problematic as the project loses their ``huge amounts of domain knowledge (P9 25m)".
The relevance of domain knowledge is also acknowledged in other work, such as attracting new project participants~\cite{Fang2023}, building effective cross-disciplinary projects~\cite{Damian2013}, and requirements engineering~\cite{Niknafs2016} (though knowing too much can sometimes be unhelpful~\cite{SHARP1991}).

Our observation of increased Scientific Debt during early software development stages reflects the challenges of integrating scientific knowledge into code. Interviews from \S \ref{sec:quaL} support this finding, as they highlight difficulties in, for example, \textit{translating mathematical algorithms into efficient Python code}, underscoring the alignment between Kelly's work, our data, and practitioner experiences.

The interviews revealed a theme surrounding complexity. As Kelly's model captures, there is essential complexity in research computing, often due to the Theory- and Real-World Knowledge domains, but not exclusively. Capturing real world phenomena (such as the presence of rocky planets in front of a distant star) is difficult and requires knowledge of astrophysics and advanced math. But this is not exclusive to these domains. Software itself has essential complexity to manage, for example, in building tools to handle exascale data volumes~\cite{ernstJSS}. Thus effective research software requires expertise across all domains. In Aranda et al.~\cite{arandaObservationsConwaysLaw2008}, the paper reports similar diffuse sets of knowledge across the team members. What seems to have changed since the Aranda paper in 2008 is that coordination mechanisms, particularly for the larger projects we study, have greatly improved communication and teamwork. All of our projects used social platforms for managing the boundary objects of source code and project management artifacts.

\subsection{Managing Research Related Technical Debt}
One constant in the TD literature is that having \emph{some} technical debt is important. If technical debt is incurred in order to learn how it should be done~\cite{cunningham_wycash_1992}, then cutting edge, novel software projects, which feature prominently in scientific research, should have a reasonable amount of technical debt (perhaps 8-15\% of the codebase~\cite{MITP_2021,Graetsch2025}).

The main question for research software projects is which types of technical debt are present.
If the technical debt is part of the discovery process inherent to research, then it is advancing project knowledge and the team's shared theory of the program (in the Naur sense of theory~\cite{Naur1985ProgrammingAT}). 
Inadvertent technical debt by contrast is undesirable, as it reflects portions of the codebase that do not expand our understanding of the program's theory, but are rather attributable to external pressures. As Ramasubbu and Kemerer note, well managed organizations understand technical debt as part of moving towards an improved end state for their products~\cite{ramasubbu}.

From our interviews, we saw that removing the less desirable technical debt is often difficult due to stakeholder focus on developing new features. This is common across software projects, and has been long-established: maintenance in general is less appealing, and often deferred~\cite{Lientz1978CharacteristicsOA}. In research software, these pressures emerge as a focus on scientific priorities (e.g., new discoveries such as pulsars), often driven by domain expert led scientific advisory committees (such as the Astropy Strategic Planning Committee\footnote{\url{https://www.astropy.org/team.html}}). This is in addition to typical funding and operational pressures (captured deductively as \textsc{cause-business-priority}, our most common deductive code). The challenge for research software projects is to allow for novel software approaches and algorithms, while ensuring that, as projects mature and grow, long-term maintainability is not sacrificed.

\subsection{Levels of Analysis and the Inevitability of Assumptions}
Scientific software must bridge multiple levels of abstraction. Marr's levels of analysis~\cite{marr1982vision}, originally proposed for understanding complex information-processing systems, provide a useful lens. At the \emph{computational level}, the software encodes what scientific problem is being solved and why (e.g., modeling ice formation in coupled climate simulations). At the \emph{algorithmic level}, it specifies how scientific theories are translated into computational procedures (e.g., choosing a truncated series expansion for the Hencky strain function). At the \emph{implementational level}, it realizes those procedures in code (e.g., choosing float vs.\ double precision for force calculations).

Our Scientific Debt indicators map naturally onto these levels. Assumptions and New Scientific Findings are primarily computational-level concerns: they reflect what the software takes to be true about the world. Translation Challenges and Computational Accuracy operate at the algorithmic level: they concern how faithfully the scientific model is rendered as a procedure. Missing Edge Cases and precision issues arise at the implementational level. Because Marr's levels are hierarchical---each level is a realization of the one above---assumptions at a higher level cascade into constraints and complications at lower levels. An assumption made to simplify the physics (computational level) forces algorithmic choices that in turn constrain the implementation, compounding debt across all three levels.

This cascade helps explain why our interviewees reported that \emph{complexity complicates}: the entanglement of scientific theory with software implementation across multiple levels makes technical debt in research software structurally different from debt in conventional systems, where the ``domain'' is typically confined to one level of abstraction. Assumptions are inevitable at every level of a complex scientific system; understanding their cross-level dependencies is key to managing Scientific Debt effectively.

\subsection{The Rise of GenAI: Technical Debt and Cognitive Debt}
\emph{Cognitive debt} is a neologism to explain the lack of understanding of a software system that arises when Generative AI tools take responsibility for a large amount of thinking (e.g., writing or programming)~\cite{cog-debt,anthropic-skill}. It refers to a poorly developed internal \emph{theory} of how the system works (or should work), as expressed by Naur~\cite{Naur1985ProgrammingAT}. Maintaining a well-grounded theory of how the system does work, and should work, is critical. From our interviews we found that this theory is developed and refined in the rich interplay between people with different knowledge domains (theory, real-world, software, execution, operations), collaborating using boundary objects such as source code and issues.  

\begin{figure}[h]
    \centering
    \includegraphics[width=0.6\columnwidth]{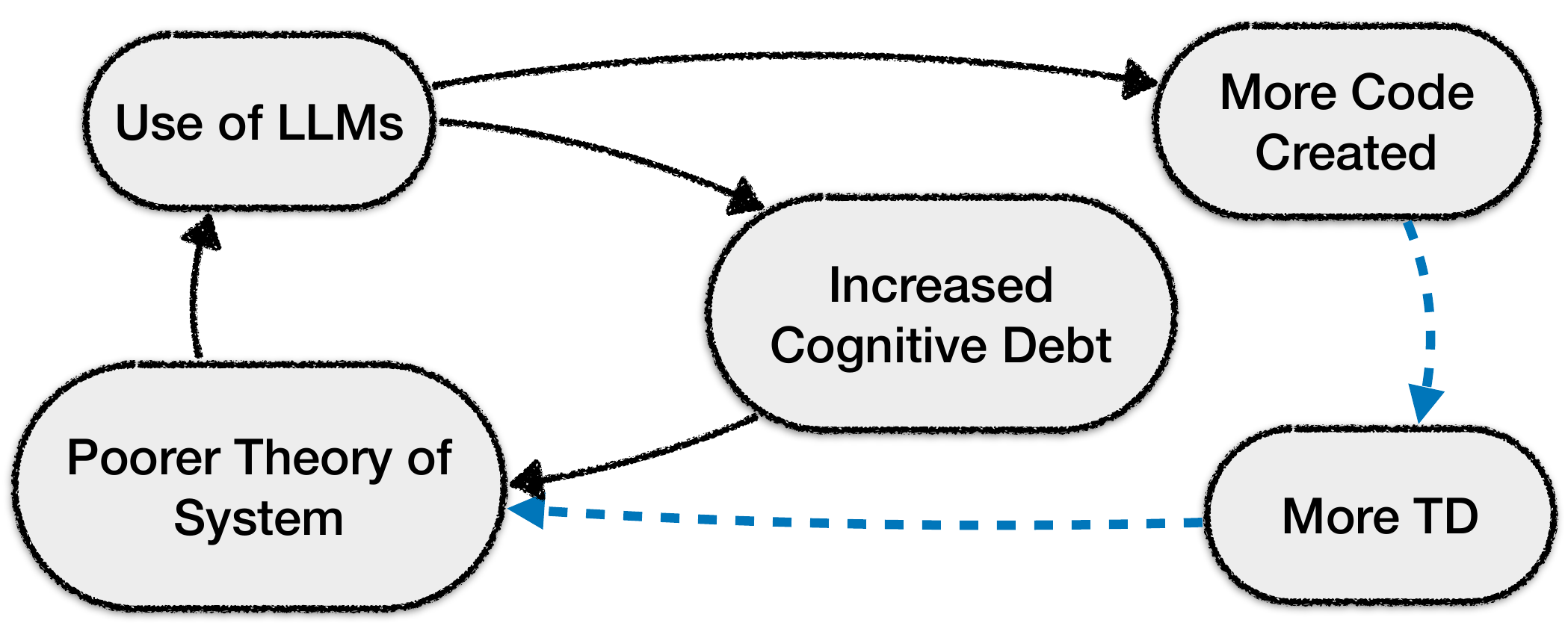}
    \caption{A possible LLM-induced cognitive/technical debt spiral.}
    \Description{Cycle diagram showing a reinforcing feedback loop: LLM use generates more code and reduces cognitive understanding, which leads to further LLM use, accelerating the accumulation of both technical and cognitive debt.}
    \label{fig:tdcd-cycle}
\end{figure}

With the rise of generative AI tools, we hypothesize---based on our findings about the importance of deep domain knowledge (Theme 3) and the cascading complexity across levels of analysis---that there is a risk of a negative feedback spiral with negative implications for long-term health of research software projects. One respondent in our member-checking survey (Section~\ref{sec:member}) independently identified LLMs as a frequent cause of accidental complexity, lending preliminary support to this concern. Figure \ref{fig:tdcd-cycle} illustrates this. Even before generative AI, as more code is written, more technical debt accrues~\cite{Curtis2012}, making the theory of the system harder to maintain (dashed lines). The system must be refactored to prevent entropy and degradation. This has been understood since the work of Lehman and Belady~\cite{Lehman1979}. In the generative AI era, this cycle accelerates (solid lines). Use of LLMs both creates more code, and reduces cognitive understanding of the system and its theory. This in turn seems likely to lead to more use of LLMs, and the cycle repeats.

Pertseva et al.~\cite{pertsevaTheoryScientificProgramming2024} also studied scientific teams. They postulated a theory of \emph{scientific programming efficacy} that suggests gradual learning curves, strong technical training, and good software engineering practice were essential to being effective in coding research software. Extending this idea of efficacy to the Naurian concept of theory~\cite{Naur1985ProgrammingAT}, generative AI tools seem to hold potential to flatten the learning curve, but at the expense of technical skills and (possibly) software engineering practices.

We emphasize that this feedback loop remains a hypothesis. Our interviews were conducted between October 2023 and June 2024, prior to widespread adoption of LLM-based coding tools in scientific computing, and participants were not asked directly about LLM usage. Systematic investigation of how generative AI affects technical debt in research software is an important direction for future work.

\subsection{Implications for Software Researchers}
While collaboration between diverse knowledge domains is crucial, the exact nature of this collaboration remains ambiguous. For instance, studies have shown that pairing scientists with software engineers can lead to challenges, as seen when a software engineer and an astronomer struggled to align unit testing with scientific goals, and another case where a software engineer faced difficulties applying standard testing practices with nuclear scientists \cite{Kelly2011ScientificST, Cote2005}. These examples suggest that simply putting different skills together may not suffice. This reinforces research results in socio-technical congruence~\cite{Damian2013,arandaObservationsConwaysLaw2008}.

While our research shows the different knowledge types necessary, it is not clear which of these is harder to acquire, or better supported in LLMs. For example, is Real-World Knowledge, which requires deep domain-specific understanding of (for example) complex mathematics and scientific principles, more challenging to gain than Software Knowledge? We found that \emph{complexity complicates}, but what type of complexity dominates? An important corollary here is that scientific projects often cannot pay skilled software professionals what they might expect, even as those individuals drive the successes of the research software projects. This results in more scientific developers starting first from scientific domains, rather than software developers acquiring science knowledge~\cite{pinto_how_2018,pertsevaTheoryScientificProgramming2024}.

Understanding these dynamics can inform the design of training programs and team structures, enhancing the effectiveness of research software development. Researchers should develop strategies that integrate diverse knowledge domains, ensuring both robustness and efficiency in research software projects.

\subsection{Implications for Research Software Projects}
The projects we studied make use of social coding platforms such as GitHub to manage code, issue tracking, and project artifacts. The code is the main source of truth, and Scientific Debt comments in the code reflect this role as boundary object between experts in the Real-World and Theory domains, and experts in Software or Operations. The Elmer project's example of the ``differential of the Hencky strain function" illustrates this, as the implementation choices and the science combine in a single shared artifact. 

The identification and categorization of Scientific Debt indicators (incorrect assumptions, missing edge cases) offer a framework for practitioners to prioritize technical debt. Understanding that new scientific findings and computational accuracy are more frequently addressed, while translation challenges and assumptions are often neglected, allows for targeted strategies. For example, a project may wish to treat novel results that ought to be incorporated into (say) a climate model differently than tradeoffs of simulation accuracy over performance.

Using LLMs like Claude Code shows promise in managing technical debt by identifying domain-specific issues. However, these tools also bring the risk of reduced understanding of the complexity inherent in research software. Integrating LLMs into the development workflow might help detect potential scientific impacts early, enabling timely interventions, although human verification, and human cognitive awareness, is necessary to improve precision and reduce false positives.

Finally, the complexity of acquiring Real-World Knowledge versus Software Knowledge suggests careful team composition. Interdisciplinary teams of scientists and software engineers can bridge knowledge gaps and enhance software reliability and accuracy. Training programs focused on both domain-specific knowledge and software engineering principles can further support collaboration.

\section{Limitations and Tradeoffs}
A mixed methods study like this one has limitations and tradeoffs from each method to report on. 

\subsection{Study 1 Limitations}
For the \textsf{quaN} study of Section \ref{sec:quaN}, we examine the internal, external, and construct validity in the study.

\textbf{Internally}, there is potential bias from the manual labeling of SATD comments. The primary categorization of 28,680 comments was conducted by the second author, with the novel category of Scientific Debt emerging during this process. Because the coder proposing the category was also the primary labeler, their evolving understanding of Scientific Debt may have shaped what was included. To mitigate this, we conducted a post-hoc inter-rater reliability check: the third author independently classified a statistically representative sample of 1,000 comments, achieving a Cohen's kappa of 0.79 (substantial agreement) computed on single-label classifications following \citet{maldonado_using_2017}. Multi-label assignments were added during subsequent disagreement resolution but were not subject to a second round of inter-rater agreement. While a concurrent independent coding process would have been stronger, the scale of the dataset (28,680 comments across multiple programming languages) made fully independent dual-coding impractical. We note that in our constructivist epistemology, novel interpretive categories are expected to emerge from close engagement with the data; the validity of Scientific Debt as a category will ultimately be confirmed or refuted through replication by subsequent studies.

Our \textbf{construct} of \emph{Scientific Debt} is novel, and raises ontological questions about how it is defined. One may ask whether such Scientific Debt comments reflect technical debt in the software engineering sense, or simply scientific uncertainty and provisional modeling. Perhaps our definition merely identified deficiencies in scientific methodology or epistemic assumptions. Our position is that these uncertainties, when embedded in the software itself---e.g., in implementing the science of `Smith and Jones 2025’---constitute technical debt because they directly affect the correctness of the software’s outputs. As Kelly~\cite{kelly_scientific_2015} writes, science, and the software implementing that science, are usually inextricable. 

The ontology and taxonomies of self-admitted technical debt are not well-defined. For example, what we call scientific technical debt might be labeled as test debt by other researchers. We acknowledge our explicit lens of scientifically relevant self-admitted technical debt. This overlap to us points to a need for a clearer ontology of terms representing these debt types.

Finally, one might ask whether Scientific Debt is unique to research software. Any software that encodes complex domain knowledge---financial modeling, medical devices, game physics---likely exhibits analogous forms of domain-knowledge debt. We would expect similar debt indicators (unvalidated assumptions, translation challenges, missing edge cases) to appear wherever software must cross boundaries between levels of abstraction, from domain theory through algorithmic realization to implementation. What distinguishes research software is the depth and explicitness of this multi-level entanglement (as discussed in Section~\ref{sec:discuss}): the explicit goal of advancing scientific understanding, the need to faithfully encode evolving theories, and the high stakes of scientific validity make the cascade of assumptions across levels particularly acute. Our definition is intentionally scoped to scientific software, but future work could investigate whether analogous categories emerge in other knowledge-intensive domains.

Our \textbf{external} validity for Study 1 is reliant on our sampling and its generalizability. 
Our sample selection was limited to projects with publicly available source code and active repositories. This focus might exclude insights from less prominent, less active, or private projects, potentially biasing our understanding toward practices in more visible and actively maintained research software.

Assessing whether Scientific Debt is a reliable signal for scientific issues within software is challenging. Our categorization aimed to capture potential scientific issues, but validating whether these comments truly reflect significant scientific concerns or are routine developer notes remains difficult. 

To address this, we conducted a thorough review process with multiple labelers and sought insights from project contributors to ensure accurate interpretations. Additionally, we cross-referenced identified Scientific Debt with actual issues and errors in the software's history to gauge the correlation between SATD and real-world scientific problems.

Finally, the reliability of our findings is influenced by the complexities of mining data from Git repositories. The decentralized nature of Git allows commits to be reordered, deleted, or edited, which can lead to inconsistencies in development history. Practices like rebasing can obscure the true sequence of events, complicating the tracking of technical debt origins and resolution. These factors necessitate cautious interpretation of our results, acknowledging potential gaps and inaccuracies in the data.

\subsection{Study 2 Limitations}
For the quaL research in Section \ref{sec:quaL}, we use the quality framework from ~\citet{small_qualitative_2022} which defines the following categories for quality and study validity: 
\emph{empathy, heterogeneity, palpability, follow-up, and self-awareness.} We explain and reflect on each in turn.
Then we explore some inherent tradeoffs we made in conducting the study.

\subsubsection{Empathy}
Empathy is about asking enough probing questions or observing enough detail to understand motivations and depth of circumstance. 

\subsubsection{Heterogeneity}
A study without heterogeneity in the findings will be a limited one that did not probe deeply. But reporting limitations often demand homogeneity to lead to actionable, generalizable insights. The suggestion in \cite{small_qualitative_2022} is to reflect carefully on the theory and use the heterogeneity to support the theory, while acknowledging that it might not cover all the variation. 

In both the quantitative and qualitative portions of the study we conducted opportunistic samples. This is partially driven by practical considerations around time and cost: scientists on these projects are usually quite busy and finding time to talk about software is difficult. But we also feel that in an exploratory study such as this one that getting a cross section of projects is more useful than a wide-net random sample that may or may not capture the interesting aspects we require (such as size, domain, openness). Another way to conceptualize this is via Yin's notion of `analytical inference'. As we select a range of cases on our dimensions of relevance, we expand the scope of potential inferences, as we may encounter a project or participant which differs from the others (e.g., as counterexample).

In particular, our interview sample skews toward climate science (6 of 11 participants), with only one participant each from astronomy, molecular dynamics, physics, and mathematics. Our themes may therefore disproportionately reflect the culture of climate modeling projects (e.g., Fortran legacy codebases, CMIP-driven release cadences, large government-funded collaborations), and may not fully capture practices in other research software domains.

A common critique of qualitative research in particular is the notion of saturation. Saturation refers to stopping conditions on the data collection process, and saturation is usually achieved only when the data analysis reveals no new insights. Our philosophical approach is at odds with this assessment. We believe it is probably impossible to saturate, particularly on rich qualitative data sources. There is still more to be learned from these interviews and research questions. What we present here is a single interpretation subject to our biases, sample bias. Further testing of the theory can only be achieved by replication.

Data saturation refers to a positivist concept of justifying when it is acceptable to "stop" the data collection process. Saturation is often measured with empirical counts of how often a given code or theme recurs, and the theory has emerged when some (typically post-hoc) threshold is met. 
In our case we stopped the interviews when the data collected was sufficient to begin analysis, and finding further interviewees was becoming difficult. 
A critique of this study therefore is that the themes that emerged below may change should we conduct further data collection. Our response is to admit that this is true, but likely true of any constructivist study. As \citet[p.201]{braunSaturateNotSaturate2021} state, it is important to be able ``to dwell with uncertainty and recognise that meaning is generated through interpretation of, not excavated from, data, and therefore judgments about `how many' data items, and when to stop data collection, are inescapably situated and subjective''

\subsubsection{Palpability} 
Palpability refers to the sense that the data are specific and concrete, acting as data points rather than generic survey questions. In this study we were careful to ask participants to reflect on a specific example of technical debt or project characteristics. The transcript reflects the specific nature of the questions and answers. At the same time we were careful not to stray too far off topic which might limit generality. 

\subsubsection{Follow-up}
\label{sec:member}
Follow-up requires the research project to validate themes and insights with the respondents and subjects of the study, as well as the broader community, including those who may not have participated in the study. 

We sent a summary of the themes and insights to our original eleven interviewees as a short survey. They were asked whether they agreed with the theme, to offer additional comments, and then asked for additional comments. Five of eleven people responded (non-response may reflect disagreement, disinterest, or simply time constraints), and all five agreed the themes represented the problem of technical debt in research software, save for one "neither agree nor disagree" for the last theme. For the open-ended answers, one comment indicated that a reason for accidental complexity may be ``uncritical adoption of practices from the software industry, whose objectives and context are very different.'' Another pointed out that new technology, in this case LLMs, was a frequent cause of accidental complexity. Finally, we posted a two page summary to the US-RSE Slack channel, a virtual watercooler for the community of research software engineers in the US, but received no feedback.

\subsubsection{Self-awareness}
More conventionally known as \emph{reflexivity}, a best practice in qualitative research is to reflect on one's positionality and bias~\cite{Sousa2025IntegratingPS}. This is because the researcher is the `data collection instrument'. We are a team of researchers in a Canadian university, in a computer science department. None of us have deep knowledge of the scientific domains we studied. The lead interviewer is a white tenured male to whom interviewees may respond in a particular way, e.g., by inferring certain influence and power relations.

Finally, we believed prior to the study that research software plays a vital role in science and society and that this role is underappreciated. 

\subsection{Tradeoffs}
Tradeoffs are inevitable in designing a study~\cite{TOSEM_2024}. Using interviews as opposed to field studies reflects our choice to focus on a detailed yet broad set of insights over realistic environments~\cite{storey_who_2020}. Interviews are usually cheaper to conduct (no field visit required) and collate (no risk of lost recordings). However, we lose insight into the non-verbal approaches our participants might reveal, for example, in how they allocate tasks or how they browse source code. 

We focus explicitly on larger code bases, that have been active for a long time, and constitute many developers. This is quite a different than other scenarios that researchers examine in research software. The other end of the scale is the small, single developer project, perhaps a PhD student thesis project. The lessons from this paper may not apply to these projects, excepting that these projects occasionally turn into bigger, more complex team projects. The German aerospace research center, DLR (Deutsche Zentrum für Luft- und Raumfahrt) categorizes research software according to \emph{application class}~\cite{hasselbringMultiDimensionalResearchSoftware2025}. This study focused on application classes 2 and 3, and not 0 or 1, which are "personal use" projects and not directly supported for external users. Navigating existing technical debt, e.g., due to changes in expected usage context, are a big part of the problem \citet{lawrenceCrossingChasmHow2018} refer to in the challenge of ``crossing the chasm" to production software projects.

\section{Conclusion}

Research software underpins modern science, yet the socio-technical challenges of its development make it fertile ground for technical debt. Using a Convergent Parallel mixed methods design, we analyzed 28,680 SATD comments across nine research software projects and interviewed 11 practitioners.

Our central contribution is \emph{Scientific Debt}, a novel category capturing the accumulation of suboptimal scientific assumptions and inaccuracies in code that can compromise the validity of results. Our interviews revealed that Scientific Debt is paradoxically driven by the same science goals that deprioritize its removal, and that managing it requires navigating the interplay of boundary objects, organizational incentives, individual expertise, and multi-layered complexity.

Together, these findings confirm that technical debt in research software is shaped by forces largely absent in commercial contexts. We contribute a new conceptual category, a labeled dataset of 28,680 comments, a reusable coding guide, and empirically grounded theory to help practitioners and funders recognize and address this challenge.

\section{CRediT Statement}
\textbf{NE}: Conceptualization, Funding acquisition, Methodology, Project administration, Supervision, Writing – original draft, Writing – review \& editing, Resources. \textbf{AA}: Data curation, Formal analysis, Investigation, Software, Visualization, Writing – original draft, Writing – review \& editing. \textbf{SH}: Writing – original draft, Writing – review \& editing, Software, Methodology. \textbf{ZL}: Methodology, Formal analysis, Data curation, Validation, Writing – original draft.

\begin{acks}
Thanks to the anonymous interview participants, and reviewers of earlier versions of this work.
This work was funded under Sloan grant G-2022-19443.

\noindent LLMs were leveraged to critique the paper, improve the presentation of plots, and with minor scripting for data analysis.

\end{acks}

\printbibliography

\end{document}